\documentclass[reqno]{amsart}
\usepackage[utf8]{inputenc}
\usepackage{amssymb, amsmath, amstext, amsopn, amsthm, amscd}
\usepackage{tikz}
\usepackage{graphicx}
\usetikzlibrary{positioning}
\usepackage[numbers]{natbib}
\usepackage{tikz-cd}

\usepackage[linktocpage=true]{hyperref}
\usepackage{algpseudocode}
\usepackage{algorithm}
\usepackage{mathtools}
\mathtoolsset{showonlyrefs=true}
\usepackage{framed}
\usepackage[disable]{todonotes}
\usepackage[margin=1em]{subfig}

\newcommand{\Hh}{\mathbb{H}}
\newcommand{\KK}{\mathfrak{k}}
\newcommand{\Rr}{\mathbb{R}}

\newcommand{\Ss}{\mathbb{S}}
\newcommand{\Tt}{\mathbb{T}}
\newcommand{\Zz}{\mathbb{Z}}
\newcommand{\GG}{\mathfrak{g}}

\newcommand{\SL}{\mathfrak{sl}}

\newcommand{\SO}{\mathfrak{so}}

\DeclareMathOperator{\Ad}{Ad}

\newcommand{\trans}{\top}

\theoremstyle{plain} 
\newtheorem{thm}{Theorem}[]

\newtheorem{lem}{Lemma}[]

\theoremstyle{definition}
\newtheorem{defn}{Definition}[]

\newtheorem{conj}{Conjecture}[]
\newtheorem{rem}{Remark}[]

\title[Integrability of point-vortex dynamics]{Integrability of point-vortex dynamics via symplectic reduction: a survey}

\date{}

\author{Klas Modin}
\author{Milo Viviani}
\address{Department of Mathematical Sciences, Chalmers University of Technology and University of Gothenburg, SE-412 96 Gothenburg, Sweden}
\email{klas.modin@chalmers.se,viviani@chalmers.se}

\begin{document}

\begin{abstract}
Point-vortex dynamics describe idealized, non-smooth solutions to the incompressible Euler equations on 2-dimensional manifolds. Integrability results for few point-vortices on various domains is a vivid topic, with many results and techniques scattered in the literature. Here we give a unified framework for proving integrability results for $N=2$, $3$, or $4$ point-vortices (and also more general Hamiltonian systems), based on symplectic reduction theory. The approach works on any 2-dimensional manifold with a symmetry group; we illustrate it on the sphere, the plane, the hyperbolic plane, and the flat torus. A systematic study of integrability is prompted by advances in 2-dimensional turbulence, bridging the long-time behaviour of 2D Euler equations with questions of point-vortex integrability. A gallery of solutions is given in the appendix.
\\[1ex]
\\[1ex]
\end{abstract}

\maketitle

\setcounter{tocdepth}{1}
\tableofcontents

\section{Introduction}

The governing equations for an incompressible inviscid fluid were formulated by Euler in 1757~~\cite{Eu1757} and have since been cardinal in the ever growing field of hydrodynamics.
On a general orientable Riemannian manifold $M$, Euler's equations are
\begin{equation}\label{eq:euler}
	\dot{\mathbf v} + \nabla_{\mathbf v}\mathbf v = -\nabla p, \qquad \operatorname{div}\mathbf v = 0
\end{equation}
where $\mathbf v$ is a vector field on $M$ describing the motion of infinitesimal fluid particles, $p$ is the pressure function, and $\nabla_\mathbf v$ denotes the co-variant derivative along $\mathbf v$.
\citet{Ar1966} discovered a rich and beautiful geometry underlying these equations; 
they can be interpreted as the geodesic equation for a right-invariant Riemannian metric on the infinite-dimensional Lie group of volume preserving diffeomorphism of $M$.
For an entry-level introduction to Arnold's discovery see the lecture notes by \citet{Mo2019}. 
More detailed expositions are given by \citet{ArKh1998} and by \citet{KhWe2009}.

101 years after Euler's influential paper on incompressible fluids, \citet{He1858} showed that the 2D Euler equations exhibit special solutions consisting of a finite number of \emph{point-vortices}. 
These are non-smooth solutions where $\operatorname{curl}\mathbf v$, called \emph{vorticity}, is a finite sum of Dirac distributions
\begin{equation}
	\omega \coloneqq \operatorname{curl}\mathbf v = \sum_{i=1}^N \Gamma_i \delta_{\mathbf r_i}.
\end{equation}
Here, $\Gamma_i \in \Rr\backslash \{0\}$ is the \emph{strength} of vortex $i$ and $\mathbf r_i \in M$ is its position ($M$ can be any 2-dimensional orientable Riemannian manifold).
If $G\colon M\times M \to \Rr$ denotes the Green's function on $M$ for the Laplacian, i.e., for $\mathbf r\in M$ we have
\begin{equation}
	\Delta G(\cdot,\mathbf r) = \delta_{\mathbf r},
\end{equation}
then the equations of motion for $N$ point-vortices are given by
\begin{equation}
	\dot{\mathbf r}_i = \sum_{i\neq j} \Gamma_j \nabla^\bot_{\mathbf r_i} G(\mathbf r_i,\mathbf r_j).
\end{equation}
Here, $\nabla^\bot_{\mathbf r_i}$ is the \emph{skew gradient} with respect to $\mathbf r_i$, i.e., the operation which first takes the gradient with respect to $\mathbf r_i$ and then rotates this vector by $\pi/2$ in positive orientation.
The point-vortex equations stated like this are rather abstract; in \autoref{sec:pvd} below we present these equations explicitly for different manifolds~$M$.

\citet{Ki1876} was the first to point out that the point-vortex system has a Hamiltonian structure; in the planar case $\mathbf r_i = (x_i,y_i)$ the canonical coordinates are given by pairs of Euclidean coordinates $x_i$ and $y_i$ scaled by the strengths $\Gamma_i$.
Thus, point-vortex dynamics fits within the realm of Hamiltonian mechanics, a central observation for understanding its phase space geometry. 


The classical way of thinking is that point-vortices constitute a formal \emph{ansatz} for solutions of the Euler equations.
By following Arnold's approach, however, \citet{MaWe1983} showed that point-vortex solutions automatically appear from a systematic study of the symmetries of the system. 

The interest in point-vortex dynamics has been growing steadily since the work of Helmholtz. 
Historically, it is fair to say that the field has emerged in two branches.
One branch, originating from \citet{On1949}, is to study a large number $N\gg 1$ of point-vortices via the tools provided by statistical mechanics for Hamiltonian dynamics.
Onsager's work laid out a statistical theory of hydrodynamics for predicting the formation of coherent structures in 2D turbulence.
Since then mathematicians and physicists have followed up on his approach, which has led to many deep and beautiful results; see \cite{MaPu2012} and references therein.\todo[color=green!40]{Add more citations here.}\
However, the question concerning the long-time behaviour of classical solutions to the 2D Euler equations is still unanswered; numerical simulations suggest a more complicated generic behaviour than predicted by theories based on statistical mechanics (see \autoref{sec:outlook} below).

The other branch is to study a few number of point-vortices, and ask whether the dynamics is integrable. 
Early contributors are \citet{Gr1877} and \citet{Po1893}, who explicitly (Gröbli) and implicitly (Poincaré) proved integrability of the planar $N=3$ case.
Since then, many results were obtained, on various domains, on integrability, equilibrium solutions, and relative equilibrium solutions; for an overview see the survey by \citet{Ha2007} and the monograph by \citet{Ne2013}.

The purpose of this paper is to advocate \emph{symplectic reduction theory} -- an underused tool in dynamical systems -- as a universal framework for proving integrability results of point-vortex dynamics (and variations thereof) on two-dimensional manifolds.
That symplectic reduction theory can be used to obtain integrability results was already stressed in the paper by \citet{MaWe1983}.
Indeed, concerning the planar case they say: ``For $N=3$ one can check that the motion is (completely) integrable in the sense that the (non-abelian) reduced phase spaces are points''.
They then go on to say: ``However one can also see that the dynamics of 3 point-vortices is (completely) integrable by exhibiting 3 independent integrals in involution\ldots''
In the literature since, mostly the second method -- to find enough first integrals in involution -- has been used.
\citet{Ki1999}, in particular, used this method to prove integrability of the $N=3$ case on any surface of constant curvature.   
However, in many ways the symplectic reduction approach is more fundamental, as it starts from only the Hamiltonian structure and the symmetries.

Concerning $N=4$, \citet{Zi1980} gave the earliest results on non-integrability. 
A few years later \citet{MaWe1983} said: ``The motion of four vortices is generally believed to be chaotic.''
Today we know that $N=4$ is a border-liner: depending on the geometry of the domain, and on other conditions such as total circulation or momentum, the $N=4$ case may be integrable.
For example, \citet{Ec1988} showed that the planar case yields integrable dynamics when restricted to the subset of solutions with vanishing total circulation and momentum.
More recently, \citet{Sa2007} showed that the $N=4$ case on the sphere with vanishing total linear momentum is integrable.
The proofs of these results are based on explicit calculations in specific coordinates.
A main point of our paper is to show that the same results quickly fall out from the symplectic reduction framework essentially without any calculations.

We remark that there is also another approach to integrability for point-vortex dynamics, based on \emph{noncommutative integrability} where the integrals of motion form a Lie algebra~\cite{BoBoMa1999,BoJo2003}.
It is possible that the symplectic reduction approach considered here and the noncommutative integrability approach are two viewpoints of the same theory.
However, such a connection is not investigated here.



Our motivation for questions of integrability of point-vortex dynamics originates from recent numerical results for 2D Euler equations indicating that integrability of point-vortex dynamics, rather than prevailing statistical mechanics based theories, is central for predicting the long-time behaviour of solutions \cite{MoVi2020}.
{}

The paper is organized as follows.
In \autoref{sec:pvd} we describe, in more detail, the point-vortex equations and their Hamiltonian structures on the sphere, the plane, the hyperbolic plane, and the flat torus.
Integrability results are given in \autoref{sec:results}.
A brief review of symplectic reduction theory is given in \autoref{sec:reduction}.
In \autoref{sec:proofs} we prove the integrability results using symplectic reduction.
A summary of the known non-integrability results are given in \autoref{sec:nonintegrability}.
Thereafter, \autoref{sec:outlook} contains an outline of the connection between long-time behaviour of 2D Euler equations and point-vortex integrability.
Finally, in \autoref{sec:gallery} we provide a gallery of point-vortex solutions -- a sort of `visual summary' of the paper.

\section{Point-vortex equations and their conservation laws}\label{sec:pvd}
In this section we give more detailed presentations of point-vortex dynamics in four specific cases: the sphere $\Ss^2$, the plane $\Rr^2$, the hyperbolic plane $\Hh^2$, and the flat torus $\Tt^2=\Rr^2/\Zz^2$.
From the point of view of symplectic reduction, these cases correspond to different structures of the symmetry group: compact non-Abelian, semi-direct product, non-compact but semi-simple, and Abelian.
These structures illustrate well the different scenarios that can occur in symplectic reduction, as we shall see in \autoref{sec:reduction} below.

\subsection{The sphere}
Consider the sphere $\Ss^2$ embedded in Euclidean 3-space as the subset of unitary vectors.
The standard area form equips $\Ss^2$ with a symplectic structure, denoted $\Omega_{\Ss^2}$.
The corresponding Poisson bracket is
\begin{equation*}
	\{F,G \}(\mathbf r) = (\nabla F \times \nabla G)\cdot \mathbf r .
\end{equation*}
We may think of $\nabla F$ and $\nabla G$ as the intrinsic Riemannian gradients on $\Ss^2$, but equally well as Euclidean gradients for extensions of $F$ and $G$ to a neighbourhood of $\Ss^2$ in $\Rr^3$; the parts of $\nabla F$ and $\nabla G$ not orthogonal to $\Ss^2$ are cancelled out in the triple product.

The phase space of $N$ point-vortices on $\Ss^2$ is given by
\begin{equation*}
	P^N_{\Ss^2} = \{ (\mathbf r_1,\ldots,\mathbf r_N) \in (\Ss^2)^N \mid \mathbf r_i \neq \mathbf r_j \text{ for } i\neq j \},
\end{equation*}
equipped with the direct product symplectic structure
\begin{equation*}
	\Omega_N(\mathbf u,\mathbf w) = \sum_{i=1}^N \Gamma_i \Omega_{\Ss^2}(\mathbf u_i,\mathbf w_i),
\end{equation*}
where, as before, $\Gamma_i$ should be interpreted as the vortex strengths and $\mathbf u$ and $\mathbf w$ are tangent vectors of $P^N_{\Ss^2}$.
The equations of motion for a Hamiltonian $H = H(\mathbf r_1,\ldots,\mathbf r_N)$ are then given by
\begin{equation}\label{eq:HamVF_sphere}
	\dot{\mathbf r}_i = \underbrace{\frac{1}{\Gamma_i}\mathbf r_i \times \nabla_{\mathbf r_i}H}_{X_H} ,
\end{equation}
where $X_H$ is called the \emph{Hamiltonian vector field}. 
Notice that, for any choice of Hamiltonian, the right hand side is tangent to $\Ss^2$ so the dynamics evolves on $(\Ss^2)^N$.
Since the Green's function on $\Ss^2$ is given by $G(\mathbf r,\mathbf r') = -\frac{1}{4\pi}\log(1-\mathbf r\cdot\mathbf r')$, the specific Hamiltonian corresponding to point-vortex dynamics is
\begin{equation}\label{eq:PV_equations_hamiltonian}
	H(\mathbf r_1,\ldots,\mathbf r_N) = -\dfrac{1}{4\pi}\sum_{i\neq j}\Gamma_i\Gamma_j\log(1-\mathbf r_i\cdot \mathbf r_j),
\end{equation}
which leads to the \emph{point-vortex equations on the sphere}
\begin{framed}
\begin{equation}\label{eq:PV_equations}
	\dot{\mathbf r}_i=\dfrac{1}{2\pi}\sum_{i\neq j}\Gamma_j\dfrac{\mathbf r_i\times \mathbf r_j}{1-\mathbf r_i\cdot \mathbf r_j}.
\end{equation}
\end{framed}

Let us now turn to the symmetries for the system \eqref{eq:PV_equations}.
Clearly, the Hamiltonian \eqref{eq:PV_equations_hamiltonian} is invariant under the diagonal action of $\mathrm{SO}(3)$ on $P^N_{\Ss^2}$.
That is, if $R\in \mathrm{SO}(3)$ then
\begin{equation*}
	H(R\mathbf r_1,\ldots,R\mathbf r_N) = H(\mathbf r_1,\ldots,\mathbf r_N).
\end{equation*}
Furthermore, this action preserves the symplectic form $\Omega_{N}$ (since it is isometric and therefore preserves the area form on each sphere).
By the Hamiltonian version of N\"other's theorem (cf.\ \cite[Thm.~11.4.1]{MaRa1999}) the symmetry, together with the fact that the action is symplectic, leads to a conservation law stated in terms of the \emph{momentum map} associated with the action of $\mathrm{SO}(3)$ on $P^N_{\Ss^2}$.
Recall that the momentum map $\mathbf J\colon P^N_{\Ss^2}\to \SO(3)^*$ is defined by the condition that, for any $\boldsymbol\xi\in\SO(3)$, the Hamiltonian vector field
\begin{equation}\label{eq:momentum_condition}
	X_{\langle \mathbf J(\cdot),\boldsymbol\xi\rangle}(\mathbf r_1,\ldots,\mathbf r_N) = \frac{1}{\Gamma_i}\mathbf r_i \times \nabla_{\mathbf r_i} \langle \mathbf J(\mathbf r_1,\ldots,\mathbf r_N),\boldsymbol\xi \rangle 
\end{equation}
corresponds to the infinitesimal action of $\xi$ on $P^N_{\Ss^2}$.
If we identify $\SO(3)$ with $\Rr^3$ via the usual \emph{hat map} (cf.~\cite[Eq.~9.2.7]{MaRa1999}), then the infinitesimal action is given by 
\begin{equation}\label{eq:inf_action_sphere}
\boldsymbol\xi\cdot (\mathbf r_1,\ldots,\mathbf r_N) = (\boldsymbol\xi\times\mathbf r_1,\ldots,\boldsymbol\xi\times\mathbf r_N),	
\end{equation}
i.e., infinitesimal rotation of each $\mathbf r_i$ about the axis $\boldsymbol\xi$. 
From \eqref{eq:HamVF_sphere} it follows that if
\begin{equation*}
	\langle \mathbf J(\mathbf r_1,\ldots,\mathbf r_N),\boldsymbol\xi\rangle = \sum_{i=1}^N\Gamma_i \mathbf r_i\cdot \boldsymbol\xi 
\end{equation*}
then the right hand side of \eqref{eq:momentum_condition} equal that of \eqref{eq:inf_action_sphere}.
Thus, identifying $\SO(3)^*$ with $\Rr^3$ via the Euclidean inner product, we get
\begin{equation}\label{eq:momentum_sphere}
	\mathbf J(\mathbf r_1,\ldots,\mathbf r_N) = \sum_{i=1}^N \Gamma_i \mathbf r_i .
\end{equation}
From N\"other's theorem it follows that the three components of $\mathbf J$, called \emph{total linear momentum}, are conserved for the point-vortex flow \eqref{eq:PV_equations}, or, more generally, for any $\mathrm{SO}(3)$-invariant Hamiltonian flow on $P^N_{\Ss^2}$.

We now come to a property of momentum maps that is central in the symplectic reduction framework, namely \emph{equivariance}.
A momentum map is called equivariant if it commutes with the symplectic action of the underlying symmetry group.
Thus, the momentum map $\mathbf J\colon P^N_{\Ss^2}\to \SO(3)^*$ is equivariant if
\begin{equation}\label{eq:equivariance_S2}
	\Ad_R^*\mathbf J(\mathbf r_1,\ldots,\mathbf r_N) = \mathbf J(R\mathbf r_1,\ldots,R\mathbf r_N) \qquad \forall\;R\in\mathrm{SO}(3),
\end{equation}
where $\Ad_R^*$ is the coadjoint action, here defined as matrix-vector multiplication. It is easy to check that this condition is fulfilled, so the momentum map $\mathbf J$ on $P^N_{\Ss^2}$ is indeed equivariant.
A general result states that if the symmetry group is semi-simple, then an equivariant momentum map always exists.
In the next section, however, we encounter a symplectic action that does not have an equivariant momentum map, which has consequences for the symplectic reduction.

\subsection{The plane}\label{sub:plane}
Consider now the Euclidean plane $\Rr^2$, with standard coordinates $\mathbf r = (x,y)$.
The canonical symplectic structure is $\Omega_{\Rr^2} \equiv dx\wedge d y$, which gives the Poisson bracket
\begin{equation*}
	\{F,G \}(\mathbf r) = \nabla F \cdot \nabla^\perp G.
\end{equation*}
where $\nabla$ is the standard gradient operator and $\nabla^\perp$ is the \emph{skew-gradient}: $\nabla^\perp G = (\partial_y G,-\partial_x G)$.
The phase space of $N$ point-vortices on $\Rr^2$ is given by
\begin{equation*}
	P^N_{\Rr^2} = \{ (\mathbf r_1,\ldots,\mathbf r_N) \in (\Rr^2)^N \mid \mathbf r_i \neq \mathbf r_j \text{ for } i\neq j \},
\end{equation*}
equipped with the scaled direct product symplectic structure
\begin{equation*}
	\Omega_N(\mathbf u,\mathbf w) = \sum_{i=1}^N \Gamma_i \Omega_{\Rr^2}(\mathbf u_i,\mathbf w_i).
\end{equation*}
The equations of motion for a Hamiltonian $H = H(\mathbf r_1,\ldots,\mathbf r_N)$ are the set of scaled canonical Hamiltonian equations
\begin{equation}\label{eq:plane}
	\dot{\mathbf r}_i = \frac{1}{\Gamma_i} \nabla_{\mathbf r_i}^\perp H .  
\end{equation}
Since the Green's function for the Laplacian on $\Rr^2$ is $G(\mathbf r,\mathbf r') = -\frac{1}{4\pi}\log(\lvert \mathbf r-\mathbf r' \rvert^2)$, the Hamiltonian for point-vortex dynamics is
\begin{equation}\label{eq:PV_equations_hamiltonian_plane}
H(\mathbf r_1,\ldots,\mathbf r_N) = -\dfrac{1}{4\pi}\sum_{i\neq j}\Gamma_i\Gamma_j \log(\lvert \mathbf r_i-\mathbf r_j\rvert^2).
\end{equation}
From \eqref{eq:plane} it follows that the corresponding \emph{point-vortex equations on the plane} are
\begin{framed}
\begin{equation}\label{eq:PV_equations_plane}
\begin{aligned}
\dot{x}_i&=-\frac{1}{2\pi}\sum_{i\neq j}\Gamma_j\frac{y_i-y_j}{\lvert \mathbf r_i-\mathbf r_j\rvert^2}\\
\dot{y}_i&=\frac{1}{2\pi}\sum_{i\neq j}\Gamma_j\frac{x_i-x_j}{\lvert \mathbf r_i-\mathbf r_j\rvert^2}.
\end{aligned}
\end{equation}
\end{framed}


Concerning the symmetries of the equations \eqref{eq:PV_equations_plane}, it is clear that the Hamiltonian \eqref{eq:PV_equations_hamiltonian_plane} is invariant to the diagonal action of the Euclidean group $\mathrm{SO}(2)\ltimes\Rr^2$. We recall that the semidirect group product is defined as:
\[
(R_1,\mathbf u_1)\cdot(R_2,\mathbf u_2) = (R_1R_2,\mathbf u_1 + R_1\mathbf u_2)
\]
for $R_1,R_2\in\mathrm{SO}(2)$ and $\mathbf u_1,\mathbf u_2\in\Rr^2$ and the action of $\mathrm{SO}(2)\ltimes\Rr^2$ on $\Rr^2$ is:
\[
(R,\mathbf u)\cdot \mathbf r = R\mathbf r + \mathbf u,
\]
for $R\in\mathrm{SO}(2)$ and $\mathbf u,\mathbf r\in\Rr^2$.
The action of the Euclidean group is clearly symplectic, being area-preserving. 
Therefore, we get the following momentum map $\textbf J:P^N_{\Rr^2}\rightarrow (\SO(2)\ltimes\Rr^2)^*$:
\begin{equation}\label{eq:PV_equations_momentum_plane}
 \mathbf J(\mathbf r_1,\ldots,\mathbf r_N) = \left(\dfrac{1}{2}\sum_{i=1}^N \Gamma_i |\mathbf r_i|^2, \sum_{i=1}^N \Gamma_i \mathbf r_i\right),
\end{equation}
whose first component is the total angular momentum, and the second component contains the total linear momenta.
However, the momentum map \eqref{eq:PV_equations_momentum_plane} is \emph{not}, in general, equivariant;
the condition for equivariance is that the total circulation vanishes, i.e., $\sum_{i}\Gamma_i = 0$. In fact, the coadjoint action of $\mathrm{SO}(2)\ltimes\Rr^2$ on $(\SO(2)\ltimes\Rr^2)^*$ is:
\[
(R,\mathbf u)\cdot(\xi,\mathbf w) = (\xi + \mathbf u^\trans(R\mathbf w),R\mathbf w)
\]
for $(R,\mathbf u)\in\mathrm{SO}(2)\ltimes\Rr^2$ and $(\xi,\mathbf w)\in(\SO(2)\ltimes\Rr^2)^*$ (see \cite[ch.~4.2]{MaMiOrRa2007}). 
Therefore, 
\[
\Ad^*_{(R,\mathbf u)} (\mathbf J(\mathbf r_1,\ldots,\mathbf r_N))=\left(\dfrac{1}{2}\sum_{i=1}^N \Gamma_i |\mathbf r_i|^2 + \mathbf u^\trans\left(\sum_{i=1}^N \Gamma_i R\mathbf r_i\right), \sum_{i=1}^N \Gamma_i R\mathbf r_i\right)
\]
and
\begin{align*}
\mathbf J((R,\mathbf u)\cdot\mathbf r_1,\ldots,(R,\mathbf u)\cdot\mathbf r_N)=&\left(\dfrac{1}{2}\sum_{i=1}^N \Gamma_i |\mathbf r_i|^2 + \mathbf u^\trans\left(\sum_{i=1}^N \Gamma_i R\mathbf r_i\right) + \dfrac{1}{2}\left(\sum_{i=1}^N \Gamma_i\right)|\mathbf u|^2,\right.\\
&\left.\sum_{i=1}^N \Gamma_i R\mathbf r_i +  \left(\sum_{i=1}^N \Gamma_i\right)\mathbf u\right)
\end{align*}
Hence, to satisfy the equivariance condition:
\begin{equation}\label{eq:equivariance_plane}
	\Ad^*_{(R,\mathbf u)} (\mathbf J(\mathbf r_1,\ldots,\mathbf r_N)) = \mathbf J((R,\mathbf u)\cdot\mathbf r_1,\ldots,(R,\mathbf u)\cdot\mathbf r_N) \qquad \forall\;(R,\mathbf u)\in\mathrm{SO}(2)\ltimes\Rr^2.
\end{equation}
we must have that the total circulation is zero, i.e., $\sum_{i}\Gamma_i = 0$.

In order to construct a momentum maps that is always equivariant, we shall explore a less obvious symmetry of equations \eqref{eq:PV_equations_plane}. Consider the group $G\subset \mathrm{GL}(2N,\Rr)$ generated by the infinitesimal generators $\xi,\eta$ defined by
\[
\xi = \left[\begin{matrix}
0 & \Gamma_1 & 0 & \Gamma_2 & \ldots & 0 & \Gamma_N \\
-\Gamma_1 & 0 &  -\Gamma_2 & 0 & \ldots & -\Gamma_N & 0\\
0 & \Gamma_1 & 0 & \Gamma_2 & \ldots & 0 & \Gamma_N \\
-\Gamma_1 & 0 &  -\Gamma_2 & 0 & \ldots & -\Gamma_N & 0\\
\vdots & \vdots & \vdots & \vdots & \vdots & \vdots & \vdots\\
0 & \Gamma_1 & 0 & \Gamma_2 & \ldots & 0 & \Gamma_N \\
-\Gamma_1 & 0 &  -\Gamma_2 & 0 & \ldots & -\Gamma_N & 0\\
\end{matrix} \right]
\hspace{.5cm}
\eta = \left[\begin{matrix}
0 & 1 & 0 & 0 & \ldots & 0 & 0 \\
-1 & 0 &  0 & 0 & \ldots & 0 & 0\\
0 & 0 & 0 & 1 & \ldots & 0 & 0 \\
0 & 0 & -1 & 0 & \ldots & 0 & 0\\
\vdots & \vdots & \vdots & \vdots & \vdots & \vdots & \vdots\\
0 & 0 & 0 & 0 & \ldots & 0 & 1 \\
0 & 0 & 0 & 0 & \ldots & -1 & 0\\
\end{matrix} \right].
\] 
We notice that $\eta$ is the infinitesimal generator of $\mathrm{SO}(2)$ and $\xi$ is the infinitesimal generator of a 1-dimensional group $K$. The symplectic form $\Omega_N$ on $\Rr^{2N}$ has the following representation in matrix form:
\[
\Omega_N = \left[\begin{matrix}
0 & \Gamma_1 & 0 & 0 & \ldots & 0 & 0 \\
-\Gamma_1 & 0 &  0 & 0 & \ldots & 0 & 0\\
0 & 0 & 0 & \Gamma_2 & \ldots & 0 & 0 \\
0 & 0 & -\Gamma_2 & 0 & \ldots & 0 & 0\\
\vdots & \vdots & \vdots & \vdots & \vdots & \vdots & \vdots\\
0 & 0 & 0 & 0 & \ldots & 0 & \Gamma_N \\
0 & 0 & 0 & 0 & \ldots & -\Gamma_N & 0\\
\end{matrix} \right].
\]
It is straightforward to check that $[\xi,\eta]=0$, $\Omega_N\xi + \xi^\trans\Omega_N=0$ and $\Omega_N\eta + \eta^\trans\Omega_N=0$. 
Therefore, $G=\mathrm{SO}(2)\times K$ is Abelian and the action of $G$ on $\Rr^{2N}$ is symplectic. 
The action of $G$ has momentum map $\mathbf L\colon P^N_{\Rr^2}\rightarrow \SO(2)^*\oplus\KK^*$:
\begin{equation}\label{eq:PV_equations_momentum_plane2}
\mathbf L(\mathbf r_1,\ldots,\mathbf r_N) = \dfrac{1}{2}\left(\sum_{i=1}^N \Gamma_i |\mathbf r_i|^2,\Big|\sum_{i=1}^N \Gamma_i \mathbf r_i\Big|^2\right).
\end{equation}
It is straightforward to check that $\mathbf L$ is an equivariant momentum map. Indeed, on the one hand $SO(2)$ acts as a diagonal isometry of $P^N_{\Rr^2}$, and therefore it preserves the Euclidean norms. On the other hand, for the action of $K$ on $\mathbf L$ we get:
\begin{align*}
\dfrac{d}{dt}\bigg|_{t=0}&\mathbf L(\exp(t\xi)(\mathbf r_1,\ldots,\mathbf r_N))=\\
&=\left(\left(\sum_{i=1}^N \Gamma_i \mathbf r_i\right)^\perp\cdot\left(\sum_{i=1}^N \Gamma_i \mathbf r_i\right),\left(\sum_{i=1}^N \Gamma_i\right)\left(\sum_{i=1}^N \Gamma_i \mathbf r_i\right)^\perp\cdot\left(\sum_{i=1}^N \Gamma_i \mathbf r_i\right)\right)=0.
\end{align*}
Hence, since $SO(2)\times K$ is Abelian, the momentum map $\mathbf L$ is equivariant.

\begin{rem}
The geometric interpretation of the one-parameter group $K$ acting on $(\mathbf r_1,\ldots,\mathbf r_N)$ is the following. 
First, if the total linear momentum $\mathbf L$ vanishes, then $K$ acts as the identity.
Second, denote by $\Gamma=\sum_i \Gamma_i$ the total circulation.
If $\mathbf L\neq 0$ then $K$ rotates synchronously the point vortices $(\mathbf r_1,\ldots,\mathbf r_N)$ in circles of radius $r = \frac{\left|\mathbf L \right|}{\Gamma}$ and with centres $\mathbf C_i$ which satisfy $\mathbf C_i + \frac{\mathbf L}{\Gamma}= \mathbf r_i$.
If the total circulation vanishes, i.e., $\Gamma=0$, then the radius becomes infinite and the motion is along straight lines.
\end{rem}

\subsection{The hyperbolic plane}
Consider the hyperbolic plane model $\Hh^2=\lbrace (x,y,z)\in\Rr^3|z^2-x^2-y^2=1,z>0\rbrace$, with Riemannian structure induced by the bilinear form on $\Rr^3$: 
\[
\mathbf a\cdot_L \mathbf b = \mathbf a\cdot (L\mathbf b),
\]
for any $\mathbf a,\mathbf b\in\Rr^3$ and $L=\mbox{diag}(-1,-1,1)$. 
As on the sphere, the volume form on $\Hh^2$ defines a symplectic structure $\Omega_{\Hh^2}$.
Defining $\mathbf a\times_L \mathbf b \coloneqq L (\mathbf a\times \mathbf b)$, the corresponding symplectic form is given by
\begin{equation*}
	\Omega_{\Hh^2}(\mathbf c)(\mathbf a,\mathbf b) = \mathbf c\cdot_L(\mathbf a\times_L \mathbf b).
\end{equation*}

The phase space of $N$ point-vortices on $\Hh^2$ is given by
\begin{equation*}
	P^N_{\Hh^2} = \{ (\mathbf r_1,\ldots,\mathbf r_N) \in (\Hh^2)^N \mid \mathbf r_i \neq \mathbf r_j \text{ for } i\neq j \},
\end{equation*}
equipped with the scalled direct product symplectic structure
\begin{equation*}
	\Omega_N(\mathbf u,\mathbf w) = \sum_{i=1}^N \Gamma_i \Omega_{\Hh^2}(\mathbf u_i,\mathbf w_i),
\end{equation*}
with $\Gamma_i$ as before and $\mathbf u$, $\mathbf w$ tangent vectors of $P^N_{\Hh^2}$.
The equations of motion for a Hamiltonian $H = H(\mathbf r_1,\ldots,\mathbf r_N)$ are 
\begin{equation}\label{eq:hyperbolic_plane}
	\dot{\mathbf r}_i = \frac{1}{\Gamma_i}\mathbf r_i\times_L \nabla_{\mathbf r_i}H .
\end{equation}
For point-vortices, the Hamiltonian is 
\begin{equation}\label{eq:PV_hyp_ham} 
H = -\frac{1}{4\pi}\sum_{i\neq j}\Gamma_i\Gamma_j \log\left(\dfrac{\mathbf r_i\cdot_L \mathbf r_j + 1}{\mathbf r_i\cdot_L \mathbf r_j - 1}\right).
\end{equation}
which gives the \emph{point-vortex equations on the hyperbolic plane} \cite{HwKi2009,HwKi2013,MoNa2014}
\begin{framed}
\begin{equation}\label{eq:PV_equations_hyp}
	\dot{\mathbf r}_i = -\frac{1}{\pi}\sum_{i\neq j}\Gamma_j \dfrac{\mathbf r_i\times_L \mathbf r_j}{(\mathbf r_i\cdot_L \mathbf r_j)^2 - 1}.
\end{equation}
\end{framed}
These equations constitute a Lie--Poisson system on $(\SL(2,\Rr)^*)^N\simeq (\Rr^3,\times_L)^N$.
Equations \eqref{eq:PV_equations_hyp} constrain the vortices to the hyperboloid $x^2+y^2-z^2=-1$. 
Furthermore, the $\mathrm{SL}(2,\Rr)$ symmetry of \eqref{eq:PV_equations_hyp} gives, analogously to $\Ss^2$, the equivariant momentum map
\begin{equation}\label{eq:PV_hyp_mom}
\mathbf J(\mathbf r_1,\ldots,\mathbf r_N) = \sum_{i=1}^N \Gamma_i \mathbf r_i.
\end{equation} 
For further background on point-vortices on the hyperbolic plane, see \cite{HwKi2009,HwKi2013,MoNa2014}.
For hydrodynamics on the hyperbolic plane, see \cite{KhMi2012}.
 
\subsection{The flat torus}
The easiest way to work with the flat torus $\Rr^2/\Zz^2$ is to extend everything to $\Rr^2$ and then assure that all operations and equations are $2\pi$-periodic in both directions.
Thus, with definitions as in \autoref{sub:plane}, the Hamiltonian for point-vortex dynamics on the flat torus, expanded to a Hamiltonian on the plane, is given by
\begin{equation}\label{eq:PV_equations_hamiltonian_torus}
	H(\mathbf r_1,\ldots,\mathbf r_N) = -\dfrac{1}{4\pi}\sum_{i\neq j}\Gamma_i\Gamma_jh(x_i-x_j,y_i-y_j),
\end{equation}
where $h(x,y) = -\dfrac{x^2}{2\pi}+\sum_{m=-\infty}^{+\infty}\log\left(\dfrac{\cosh(x-2\pi m)-\cos(y)}{\cosh(2\pi m)}\right)$.
The corresponding equations of motion (cf.~\cite{WeMcM1991}) are
\begin{framed}
\begin{equation}\label{eq:PV_equations_torus}
\begin{aligned}
&\dot{x}_i=-\frac{1}{2\pi}\sum_{i\neq j}\sum_{m=-\infty}^{+\infty}\Gamma_j\frac{\sin(y_i-y_j)}{\cosh(x_i-x_j-2\pi m)-\cos(y_i-y_j)}\\
&\dot{y}_i=\frac{1}{2\pi}\sum_{i\neq j}\sum_{m=-\infty}^{+\infty}\Gamma_j\frac{\sinh(x_i-x_j)}{\cosh(y_i-y_j-2\pi m)-\cos(x_i-x_j)}-\frac{x_i-x_j}{\pi}.
\end{aligned}
\end{equation}
\end{framed}
These equations are (equivalent to) the \textit{point-vortex equation on a flat torus}, as explained in the following remark. 

\begin{rem}\label{rem:rem_1}
The equations \eqref{eq:PV_equations_torus} can be derived in two equivalent ways. 
The first, proposed in \citep{WeMcM1991}, starts from a $2\pi$-periodic distribution of vortices in the point-vortex equation on $\Rr^2$ and sums up the infinite number of contributions given by the different vortices, obtaining in the limit a well defined vector field for the right hand side in \eqref{eq:PV_equations_torus}. 
As shown in \citep{WeMcM1991}, this vector field is $2\pi$-periodic. 
Therefore it can be seen as a vector field on $\Tt^2$. 
The second way consist of explicitly calculate the Green's function of the Laplacian on a flat 2-torus. 
The calculations given in \citep{Mam2014} confirm the equivalence of the two approaches, an the description of the Green's function in terms of the First Jacobi Theta function guarantees the smoothness of the vector field on $P^N_{\Rr^2}$.
We stress that this equivalence is crucial in the analysis of the equations \eqref{eq:PV_equations_torus}. 
In fact, on the one hand we have that the 2-torus has non-trivial first de Rham cohomology, whereas on the other hand for $\Rr^2$ the cohomology is trivial, and so for any infinitesimal symmetry we get a conservation law. 
In particular, we get two more first integrals as stated below.
\end{rem}

The Hamiltonian \eqref{eq:PV_equations_hamiltonian_torus} is invariant with respect to the diagonal action of $\Rr^2$. Since this action is symplectic, we get the (linear) momentum map
\begin{equation}\label{eq:PV_equations_momentum_torus}
\mathbf J(\mathbf r_1,\ldots,\mathbf r_N) = \sum_{i=1}^N \Gamma_i \mathbf r_i.
\end{equation}
As noticed in remark~\ref{rem:rem_1}, it is convenient to consider the point-vortex equations in $\Rr^2$ according to the Hamiltonian \eqref{eq:PV_equations_hamiltonian_torus}. 
In this setting, the $\Rr^2$ action on $\Rr^2$ is free and proper, being the map $R$ in definition~\ref{def:prop_act} invertible. 
However, the momentum map \eqref{eq:PV_equations_momentum_torus} is in general \emph{not} equivariant. 
In fact, since $\Rr^2$ is an Abelian group, the $\Ad$ operator is the identity. 
Hence, the equivariance equation $\Ad^*_{g^{-1}}\circ\mu=\mu\circ\rho_g$ implies that $\mu=\mu\circ\rho_g$. 
We notice that this is true only if the circulation is zero, i.e., $\Gamma_1+\dots+\Gamma_N=0$. 
Moreover, since $\Rr^2$ is neither compact nor semisimple, Theorem~\citep[Thm 11.5.2]{MaRa1999} cannot be applied to get a modified equivariant momentum map. 
We notice that considering the point-vortex equation directly on the flat 2-torus and the action of $U(1)^2$ instead of $\Rr^2$ would have in principle guarantee the equivariance of the momentum map, but the non-trivial first group of De Rham cohomology of the 2-torus would have prevent the existence of the momentum map itself.

\section{Integrability results} \label{sec:results}

In this section we state results on integrability for the point-vortex equations presented in the previous section.
As we shall see in the sections below, all of these results, which are valid not just for the point-vortex equations but for all invariant Hamiltonian systems on the respective phase spaces, follow directly from the symplectic reduction framework.
Most of the results are known since before; the point here is to demonstrate how the results naturally fall out from symplectic reduction in a purely geometric way.
In particular, symplectic reduction gives a clear geometric understanding for why the vanishing momentum and circulation cases are structurally different from the non-vanishing cases.

Before proceeding to the specific manifolds $\Ss^2$, $\Rr^2$, $\Hh^2$, and $\Tt^2$, we recall the following general definition:

\begin{defn}
	Let $P$ be a phase space manifold for a dynamical system and assume that $P$ is acted upon by a Lie group $G$.
	A solution $t\mapsto \mathbf r(t) \in P$ is called a $G$-\emph{relative equilibrium} if there exists an $\mathbf r_0\in P$ and $\boldsymbol\xi\in\mathfrak{g}$ such that $\mathbf r(t) = \exp(t\boldsymbol\xi)\cdot \mathbf r_0$. 
\end{defn}

\subsection{The sphere}
Integrability of the point-vortex equations~\eqref{eq:PV_equations} for $N=3$ is given by \citet{Sa1999}.
The proof is based on specific coordinates for which one finds three first integrals in involution.
Eight years later \citet{Sa2007} also proved integrability of the vanishing momentum $N=4$ case, essentially by reducing it to the $N=3$ case.
See also the work by \citet{BoKiMa2005}.

\begin{thm}[$\Ss^2$ with non-vanishing momentum]\label{thm:PV_int_S2_non_zero_mom}
Consider Hamilton's equations~\eqref{eq:HamVF_sphere} on $P_{\Ss^2}^N$ for an $\mathrm{SO}(3)$ invariant Hamiltonian, restricted to the part of phase space with non-vanishing total linear momentum
\begin{equation*}
	\{(\mathbf r_1,\ldots,\mathbf r_N) \in P^N_{\Ss^2}\mid \mathbf J(\mathbf r_1,\ldots,\mathbf r_N) \neq \mathbf 0 \}.
\end{equation*}
\begin{itemize}
	\item If $N=2$ all solution are relative equilibria for the $\mathrm{SO}(2)$-action, the isotropy group of the momentum $\mathbf J$.
	\item If $N=3$ the system is completely integrable (but solutions are typically not relative equilibria). 
\end{itemize}
In particular, these results are valid for the point-vortex equations \eqref{eq:PV_equations} on $P_{\Ss^2}^N$.
\end{thm}

\begin{thm}[$\Ss^2$ with vanishing momentum]\label{thm:PV_int_S2_zero_mom}
Consider Hamilton's equations~\eqref{eq:HamVF_sphere} on $P_{\Ss^2}^N$ for an $\mathrm{SO}(3)$ invariant Hamiltonian, restricted to the part of phase space with vanishing total linear momentum
\begin{equation}\label{eq:zero_mom}
	\{(\mathbf r_1,\ldots,\mathbf r_N) \in P^N_{\Ss^2}\mid \mathbf J(\mathbf r_1,\ldots,\mathbf r_N) = \mathbf 0 \}.
\end{equation}
\begin{itemize}
\item If $N=2$ all solutions are equilibria of antipodal points.
\item If $N=3$ all solution are relative equilibria for the $\mathrm{SO}(3)$-action.
\item If $N=4$ the system is completely integrable (but solutions are typically not relative equilibria).
\end{itemize}
In particular, these results are valid for the point-vortex equations \eqref{eq:PV_equations} on $P_{\Ss^2}^N$.
\end{thm}

\subsection{The plane}
Integrability of point-vortices in the plane were the first to be studied.
Early results for $N=3$ were given by \citet{Gr1877} and by \citet{Po1893}.
The $N=4$ result was given by \citet{Ec1988}.

\begin{thm}[$\Rr^2$ with non-vanishing circulation or momentum]\label{thm:PV_int_plane_non_zero_circ_zero_mom}
Consider Hamilton's equations~\eqref{eq:plane} on $P_{\Rr^2}^N$ for an $\mathrm{SO}(2)\times K$ invariant Hamiltonian, restricted to the case of non-vanishing circulation
\begin{equation}
\begin{array}{ll}
\sum_{i=1}^N\Gamma_i \neq 0\\
\end{array}
\end{equation}
or restricted to the part of phase space with non-vanishing total linear momentum
\begin{equation}
\begin{array}{ll}
\sum_{i=1}^N \Gamma_i \mathbf r_i \neq 0.
\end{array}
\end{equation}
\begin{itemize}
\item If $N=2$ and $\sum_i \Gamma_i \neq 0$ (non-vanishing circulation), then the solutions are $\mathrm{SO}(2)\times K$ relative equilibria.
\item If $N=2$ and $\sum_i \Gamma_i = 0$ (vanishing circulation), then the solutions are $\Rr^2$-relative equilibria (travelling vortex dipoles).
\item If $N=3$ the system is completely integrable (but solutions are typically not relative equilibria).
\end{itemize}
In particular, these results are valid for the point-vortex equations \eqref{eq:PV_equations_plane} on $P_{\Rr^2}^N$.
\end{thm}

\begin{thm}[$\Rr^2$ with vanishing circulation and momentum]\label{thm:PV_int_plane_zero_circ_zero_mom}
Consider Hamilton's equations~\eqref{eq:plane} on $P_{\Rr^2}^N$ for an $\mathrm{SO}(2)\ltimes \Rr^2$ invariant Hamiltonian, restricted to case of vanishing circulation and total linear momentum:
\begin{equation}\label{eq:zero_circ}
\sum_{i=1}^N\Gamma_i = 0, \quad \sum_{i=1}^N \Gamma_i \mathbf r_i = 0.
\end{equation}
\begin{itemize}
\item If $N=2$, the point-vortex equation \eqref{eq:PV_equations_plane} is not defined.
\item If $N=3$ all solutions are $\mathrm{SO}(2)\ltimes \Rr^2$ relative equilibria.
\item If $N=4$ the system is completely integrable (but solutions are typically not relative equilibria).
\end{itemize}
In particular, these results are valid for the point-vortex equations \eqref{eq:PV_equations_plane} on $P_{\Rr^2}^N$.
\end{thm}

\subsection{The hyperbolic plane}

Integrability results on the hyperbolic plane reflect the results on the sphere, as from a symplectic reduction point of view, the two settings are almost the same, with an equivariant momentum map for a semi-simple 3-dimensional symmetry group.

\begin{thm}[$\Hh^2$ with non-vanishing momentum]\label{thm:PV_int_hyp_non_zero_mom}
Consider Hamilton's equations~\eqref{eq:hyperbolic_plane} on $P_{\Ss^2}^N$ for an $\mathrm{SL}(2)$ invariant Hamiltonian, restricted to the part of phase space with non-vanishing total linear momentum
\begin{equation*}
	\{(\mathbf r_1,\ldots,\mathbf r_N) \in P^N_{\Hh^2}\mid \mathbf J(\mathbf r_1,\ldots,\mathbf r_N) \neq \mathbf 0 \}.
\end{equation*}
\begin{itemize}
	\item If $N=2$ all solution are relative equilibria for the action of the isotropy subgroup of the momentum $\mathbf J$.
	\item If $N=3$ the system is completely integrable (but solutions are typically not relative equilibria). 
\end{itemize}
In particular, these results are valid for the point-vortex equations \eqref{eq:PV_equations_hyp} on $P_{\Hh^2}^N$.
\end{thm}

\begin{thm}[$\Hh^2$ with vanishing momentum]\label{thm:PV_int_hyp_zero_mom}
Consider the point-vortex equation \eqref{eq:PV_equations_hyp} restricted to the part of phase space with vanishing total linear momentum
\begin{equation*}
	\{(\mathbf r_1,\ldots,\mathbf r_N) \in P^N_{\Hh^2}\mid \mathbf J(\mathbf r_1,\ldots,\mathbf r_N) = \mathbf 0 \}.
\end{equation*}
\begin{itemize}
\item If $N=2$ the point-vortex equation \eqref{eq:PV_equations_hyp} is not defined.
\item If $N=3$ all solution are relative equilibria for the $\mathrm{SL}(2)$-action.
\item If $N=4$ the system is completely integrable (but solutions are typically not relative equilibria).
\end{itemize}
In particular, these results are valid for the point-vortex equations \eqref{eq:PV_equations_hyp} on $P_{\Hh^2}^N$.
\end{thm}

\subsection{The flat torus}
Integrability results for point-vortex dynamics on the flat torus were given by \citet{AreStr1999}, in the case of zero total circulation, and \citet{KiAr2018}.
As they do, we take here the view-point that the phase space is embedded as periodic solutions in $P_{\Rr^2}^N$.

\begin{thm}[$\Tt^2$ with non-vanishing circulation]\label{thm:PV_int_T2_non_zero_circ}
Consider Hamilton's equations~\eqref{eq:plane} on $P_{\Rr^2}^N$ for an $\Rr^2$ invariant Hamiltonian, restricted to the case of non-vanishing circulation:
\begin{equation} 
\sum_{i=1}^N\Gamma_i \neq 0.
\end{equation}
\begin{itemize}
\item If $N=2$ the system is completely integrable (but solutions are typically not relative equilibria).
\end{itemize}
In particular, these results are valid for the point-vortex equations \eqref{eq:PV_equations_torus} on $P_{\Tt^2}^N$.
\end{thm}

\begin{thm}[$\Tt^2$ with vanishing circulation]\label{thm:PV_int_T2_zero_circ}
Consider Hamilton's equations~\eqref{eq:plane} on $P_{\Rr^2}^N$ for an $\Rr^2$ invariant Hamiltonian, restricted to the case of vanishing circulation
\begin{equation}
\sum_{i=1}^N\Gamma_i = 0.
\end{equation}
\begin{itemize}
\item If $N=2$ all solutions are $\Rr^2$-relative equilibria.
\item If $N=3$ the system is completely integrable (but solutions are typically not relative equilibria).
\end{itemize}
In particular, these results are valid for the point-vortex equations \eqref{eq:PV_equations_torus} on $P_{\Tt^2}^N$.
\end{thm}

\section{Symplectic reduction theory}\label{sec:reduction}
In order to prove the theorems stated in section~3, we need to recall some definitions and notations. We will denote in the following the smooth action of a Lie group $G$ on a manifold $M$ with:
\begin{align*}
&\rho : G\times M\rightarrow M\\
&\rho(g,p)=:\rho_g (p)=:g\cdot p,
\end{align*} 
for each $g\in G,p\in M$.
\begin{defn}[Symplectic action]
Let $G$ be a Lie group acting on a smooth symplectic manifold $(M,\omega)$. Then the action of $G$ is said to be symplectic if $\rho^*_g\omega = \omega$, for every $g\in G$.
\end{defn}
\begin{defn}[Momentum map]
Let $G$ be a Lie group and $\GG$ its Lie algebra. Assume that $G$ acts symplectically on a smooth symplectic manifold $(M,\omega)$ and that $\xi_p:=d_e\rho_{\exp(\xi)}(p)$, for every $p\in M$, is its inifinitesimal action. Then a map $\mu:M\rightarrow \GG^*$, defined by:
\[
d\langle\mu(p),\xi\rangle=\iota_{\xi_p}\omega,
\]
is called momentum map. 
\end{defn}
When $G$ is compact,  and a momentum map exists, it can always be chosen to satisfy the equivariance identity:
\[
\langle\mu(p),[\xi,\eta]\rangle=\lbrace\langle\mu(p),\xi\rangle,\langle\mu(p),\eta\rangle\rbrace,
\]
for $\xi,\eta\in\GG$ and $p\in M$, where the bracket $\lbrace\cdot,\cdot\rbrace$ is the Poisson bracket on $M$. In this case the Lie group action will be said to be \textit{Hamiltonian}.
A crucial fact on momentum maps is that they are conserved quantities for Hamiltonian systems with symmetries.
\begin{thm}[\citep{MaRa1999}]
Let $M$ be a smooth symplectic manifold and let $G$ be a Lie group with Hamiltonian action on it and momentum map $\mu:M\rightarrow\GG^*$. Let $H:M\rightarrow\Rr$ be a smooth function such that, for each $g\in G,x\in M$, $H(g\cdot x)=H(x)$. Then, if $\phi^H_t$ is the flow of the Hamiltonian vector field $X_H$, we have that $(\phi^H_t)^*\mu=\mu$, for each $t\geq 0$.
\end{thm}
Let us now continue defining some properties of the group actions.
\begin{defn}[Free action]
Let $G$ be a Lie group acting on a smooth manifold $M$. Then the action of $G$ is said to be free if, for each $x\in M$, $g\cdot x=x$ implies $g=e$.
\end{defn}
\begin{defn}[Proper action]\label{def:prop_act}
Let $G$ be a Lie group acting on a smooth manifold $M$. Then the action of $G$ is said to be proper if the map $R :G\times M\rightarrow M\times M$, $R(g,p)=(\rho_g(p),p)$ is a proper map of topological spaces.
\end{defn}
\begin{defn}[Isotropy subgroup]
Let $G$ be a Lie group acting on a smooth manifold $M$. Then the set:
\[
G_p = \lbrace g\in G \mid g\cdot p = p\rbrace
\]
is a Lie subgroup of $G$ called isotropy subgroup.
\end{defn}
Let us now recall the symplectic reduction theorem:
\begin{thm}[Symplectic reduction theorem, cf.~\cite{MaMiOrRa2007}]\label{thm:Symplectic Reduction theorem}
Let $G$ be a Lie group with Hamiltonian free and proper action on a smooth symplectic manifold $(M,\omega)$ and let $\mu:M\rightarrow \GG^*$ be an equivariant momentum map with respect to this action. Then, for each $\phi\in\GG^*$, the quotient:
\[
J_\phi:=\mu^{-1}(\phi)/G_\phi
\]
is a symplectic manifold of dimension $d:=\dim(M)-2\dim(G_\phi)$, with symplectic form $\omega_\phi$ uniquely characterized by 
\[\pi^*\omega_\phi=\iota^*\omega,\]
 where $\pi:\mu^{-1}(\phi)\rightarrow \mu^{-1}(\phi)/G_\phi$ is the projection and $\iota: \mu^{-1}(\phi)\rightarrow M$ is the inclusion.
\end{thm}

\section{Proofs by symplectic reduction}\label{sec:proofs}

In this section we prove the results stated in \autoref{sec:results}.
All of our proofs are based on the symplectic reduction Theorem~\ref{thm:Symplectic Reduction theorem}, which makes them very streamlined.
Essentially, the proofs consist in checking that the action is free and proper, that the momentum map is equivariant, and then counting the dimensions of the reduced phase space.

\subsection{The sphere}
For the sphere, the $\mathrm{SO}(3)$ action on $P_{\Ss^2}^N$ is free for any $N\geq 3$ and for $N=2$ if the points are not antipodal.
It is also proper since $\mathrm{SO}(3)$ is compact. 
Furthermore, the momentum map \eqref{eq:momentum_sphere} is equivariant, as one can directly check. 

\proof[Proof of Theorem~\ref{thm:PV_int_S2_non_zero_mom}] For $N=2$, the $\mathrm{SO}(3)$ action is free, unless the two vortices are antipodal, which from \eqref{eq:PV_equations} implies that solutions are equilibria. 
Hence, Theorem~\ref{thm:Symplectic Reduction theorem} tells us that systems evolves on a zero-dimensional manifold, since $G_\phi\simeq \mathrm{SO}(2)$. 
Thus, the reconstructed motion is an $\mathrm{SO}(3)$ relative equilibrium, i.e., a steady rotation of the initial positions.

For $N=3$, the $\mathrm{SO}(3)$ action is always free and Theorem~\ref{thm:Symplectic Reduction theorem} tells us that the system evolves on a 2-dimensional manifold. 
Any Hamiltonian system on a 2-dimensional manifold is integrable, so the reconstructed system is also integrable, as follows, for example, from the standard Floquet theory (cf.~\cite[Proof of thm~3.1]{MoVe2014}).

\endproof

\proof[Proof of Theorem~\ref{thm:PV_int_S2_zero_mom}] For $N=2$, the vanishing momentum condition \eqref{eq:zero_mom} implies that $\mathbf r_1,\mathbf r_2$ are antipodal points with equal strength. 
From \eqref{eq:PV_equations} it follows directly that solutions are equilibria.

For $N=3$, the $\mathrm{SO}(3)$ action is free and Theorem~\ref{thm:Symplectic Reduction theorem} tells us that the reduced system evolves on a zero-dimensional manifold, so solutions are relative equilibria.

For $N=4$, the $\mathrm{SO}(3)$ action is free and by Theorem~\ref{thm:Symplectic Reduction theorem} the system evolves on a 2-dimensional reduced manifold, so the dynamics is integrable. 
\endproof
%

\subsection{The plane}
The action of $\mathrm{SO}(2)$ on $P_{\Rr^2}^N$ is free for $N=1$ unless $\mathbf r_1 =\mathbf 0$.
The action is always free for $N\geq 2$. The action of $K$ is free and proper only under certain conditions, as stated in the following lemmas.

\begin{lem}\label{lem:plane_1}
If $\sum\Gamma_i\mathbf r_i\neq 0$, the action $\rho:K\times P_{\Rr^2}^N\rightarrow P_{\Rr^2}^N$ is free.
\end{lem}
\proof Since $K$ is 1-dimensional, the action of $K$ is non-free if and only if $K$ has fixed points. Let $\xi$ be the infinitesimal generator of $K$, as defined in section~\ref{sub:plane}. Then the action $\rho$ of $K$ is free if and only if the kernel of $\xi$ is trivial. 
It is straightforward to check that the kernel of $\xi$ is given by:
\[
\mbox{ker}(\xi)=\lbrace (\mathbf r_1,\dots,\mathbf r_N)\in P_{\Rr^2}^N | \sum\Gamma_i\mathbf r_i = 0 \rbrace.
\]
Hence, under the non-vanishing linear momentum condition the action of $K$ is free at any time.
\endproof
Since the linear momentum is a first integral of the equations~\eqref{eq:PV_equations_hamiltonian_plane}, it follows from Lemma~\ref{lem:plane_1} that the $K$ action is free provided that the linear momentum is non-vanishing for the initial condition. 

\begin{lem}\label{lem:plane_2}
If $\sum\Gamma_i\neq 0$, the action $\rho:K\times P_{\Rr^2}^N\rightarrow P_{\Rr^2}^N$ is proper.
\end{lem}
\proof Let $\xi$ be the infinitesimal generator of $K$, as defined in section~\ref{sub:plane}. Then, it is straightforward to check that $\xi$ has rank 2, for any $N\geq 1$ and its non-zero eigenvalues are purely imaginary and equal to $\pm i\sum\Gamma_i\neq0$. Hence, the group $K$ is bounded and closed in the operator norm topology. Therefore, it is compact and so its action is proper.
\endproof
For $N\geq 2$, the action of $\mathrm{SO}(2)\ltimes \Rr^2$ on $P_{\Rr^2}^N$ is free and proper. Indeed, if $R\in SO(2)$ is not the identity, $v\in\Rr^2$ and $(p,q)\in P_{\Rr^2}^2$, then $Rp+v=p$ and $Rq+v=q$ imply $p=q$, since $R$ does not have real eigenvectors, and so the action is free. Moreover, $SO(2)$ is compact and the map $R$ in definition~\ref{def:prop_act} for $\Rr^2$-action has continuous inverse, therefore the semidirect product action is a composition of proper maps and so it is proper.

\proof[Proof of Theorem~\ref{thm:PV_int_plane_non_zero_circ_zero_mom}] 
We divide the proof in two cases.

\textbf{Case 1:} $\sum_i\Gamma_i\neq 0$. Without loss of generality due to the translational invariance of equations \eqref{eq:PV_equations_hamiltonian_plane} and to the non-vanishing total circulation, we can assume that the linear momentum is non-zero. Then, lemmas~\ref{lem:plane_1}-\ref{lem:plane_2} ensure the $\mathrm{SO}(2)\times K$ action is free and proper, for any $N\geq 2$.

For $N=2$, Theorem~\ref{thm:Symplectic Reduction theorem} tells us that the reduced Hamiltonian system has dimension 0, and therefore the motion is a $\mathrm{SO}(2)\times K$-relative equilibrium. 

For $N=3$, Theorem~\ref{thm:Symplectic Reduction theorem} with respect to the $\mathrm{SO}(2)\times K$ action tells us that the reduced Hamiltonian system has dimension $2$, which implies that it is integrable. 

\textbf{Case 2:} $\sum_i\Gamma_i = 0$ If the circulation is zero, the linear momentum map due to the action of $\Rr^2$ is equivariant. Moreover, the $\Rr^2$ action is always free and proper for $N\geq 1$.

For $N=2$, Theorem~\ref{thm:Symplectic Reduction theorem} tells us that the reduced Hamiltonian system has dimension 0 when $N=2$. Therefore the motion can only be up to translations; this explains the vortex dipole solutions.

For $N=3$, Theorem~\ref{thm:Symplectic Reduction theorem} tells us that the reduced Hamiltonian system has dimension $2$, which implies that it is integrable. 
\endproof

\proof[Proof of Theorem~\ref{thm:PV_int_plane_zero_circ_zero_mom}] For $N=2$, the zero-circulation and zero-momentum conditions imply that $\mathbf r_1=\mathbf r_2$, so the point-vortex equation \ref{eq:PV_equations_plane} is not defined.

For $N=3$, the $\mathrm{SO}(2)\ltimes\Rr^2$ the momentum map is equivariant, so Theorem~\ref{thm:Symplectic Reduction theorem} tells us that the reduced systems evolves on a zero-dimensional manifold. 
Therefore, the motion of the point-vortices is a $\mathrm{SO}(2)\ltimes\Rr^2$ relative equilibrium.

For $N=4$, the $\mathrm{SO}(2)\ltimes\Rr^2$ the momentum map is equivariant, so Theorem~\ref{thm:Symplectic Reduction theorem} tells us that the reduced systems evolves on a on a 2-dimensional manifold. 
Integrability then follows. 
\endproof

\subsection{The hyperbolic plane}
In order to get a free action for $N\geq 2$, we need to restrict to $\mathrm{PSL}(2)\cong \mathrm{SL}(2)/\lbrace\pm 1\rbrace$. Furthermore, it is known that the action is proper \citep{MoNa2014}.

\proof[Proof of Theorem~\ref{thm:PV_int_hyp_non_zero_mom}] For $N=2$, Theorem~\ref{thm:Symplectic Reduction theorem} tells us that systems evolves on a zero-dimensional manifold, since $G_\phi$ is a 1-dimensional Lie group (see~\citep{MoNa2014}). 
Thus, solutions are relative equilibria with respect to Möbius transformations.

For $N=3$, Theorem~\ref{thm:Symplectic Reduction theorem} tells us that systems evolves on a 2-dimensional manifold, so the dynamics is integrable. 
\endproof

\proof[Proof of Theorem~\ref{thm:PV_int_hyp_zero_mom}] For $N=2$, the zero momentum condition \eqref{eq:zero_mom} implies that $\mathbf r_1,\mathbf r_2$ lie on common line through the origin. 
Hence, $\mathbf r_1=\mathbf r_2$ and so the equations \eqref{eq:PV_equations_hyp} are not defined.

For $N=3$, the $\mathrm{PSL}(2)$ action is free and Theorem~\ref{thm:Symplectic Reduction theorem} tells us that systems evolves on a zero-dimensional manifold. 
Thus, solutions are relative equilibria with respect to Möbius transformations.

For $N=4$, the $\mathrm{PSL}(2)$ action is free and Theorem~\ref{thm:Symplectic Reduction theorem} tells us that systems evolves on a 2-dimensional manifold, so the dynamics is integrable. 
\endproof

\subsection{The flat torus}
The action of $\Rr^2$ on $\Tt^2$ is not free, but as we have seen in \autoref{sec:pvd}, we understand the point-vortex equations on $\Tt^2$ as a special case of the point-vortex dynamics on $\Rr^2$. 
Therefore, the action of $\Rr^2$ on itself via translations is free and proper.
We can also consider the action of the group $K$, as defined in section~\ref{sub:plane}. Since its momentum map only depends on the conservation of the $\Rr^2$ momentum map, $K$ is a symmetry also of any translational invariant Hamiltonian on the torus.

\proof[Proof of Theorem~\ref{thm:PV_int_T2_non_zero_circ}] 
For $N=2$, without loss of generality due to the translational invariance of equations \eqref{eq:PV_equations_hamiltonian_torus} and to the non-vanishing total circulation, we can assume that the linear momentum is non-zero. Then, lemmas~\ref{lem:plane_1}-\ref{lem:plane_2} ensure the $K$ action is free and proper. Therefore, Theorem~\ref{thm:Symplectic Reduction theorem} tells us that systems evolves on a 2-dimensional manifold, so the dynamics is integrable. 
\endproof

\proof[Proof of Theorem~\ref{thm:PV_int_T2_zero_circ}] For $N=2$, the $\Rr^2$ action is free and Theorem~\ref{thm:Symplectic Reduction theorem} tells us that systems evolves on a zero-dimensional manifold. This means that the initial condition can only be transformed via translations, which conserve the Hamiltonian.

For $N=3$, the $\Rr^2$ action is free and Theorem~\ref{thm:Symplectic Reduction theorem} tells us that systems evolves on a 2-dimensional manifold. 
By definition, any Hamiltonian system on a 2-dimensional manifold is integrable.
\endproof

\section{Non-integrability results}\label{sec:nonintegrability}
In this section we briefly summarize the results in literature on non-integrability of point-vortex dynamics. 
Unlike the integrability results, which have been addressed extensively, only the planar case has been fully analysed and completed. 
Some results concern non-integrability for a restricted model of the point-vortex dynamics, defined as follows:

\begin{defn} The \emph{restricted problem of $N$ point-vortices} consists of a partition of the vortices in two sets $A$ and $B$, such that the vortices in $A$ have non-zero strength and do not interact with vortices in $B$, and the vortices in $B$ have vorticity $0\leq \varepsilon \ll 1$ and interact with all the other vortices.
\end{defn}

\begin{thm}[$\Ss^2$ with non-vanishing momentum, \cite{BagBag1997}]
Consider the point-vortex equations~\eqref{eq:PV_equations} on $P_{\Ss^2}^N$, restricted to the part of phase space with non-vanishing total linear momentum
\begin{equation*}
	\{(\mathbf r_1,\ldots,\mathbf r_N) \in P^N_{\Ss^2}\mid \mathbf J(\mathbf r_1,\ldots,\mathbf r_N) \neq \mathbf 0 \}.
\end{equation*}
\begin{itemize}
	\item If $N=4$, the restricted model with $A$ of three identical vortices and $B$ of one single vortex with strength $\varepsilon=0$ is non-integrable. 
\end{itemize}
\end{thm}

\begin{thm}[$\Rr^2$ with non-vanishing circulation or momentum, \cite{Zi1980,Zi1982,KoiCar1989}]
Consider Hamilton's equations~\eqref{eq:PV_equations_plane} on $P_{\Rr^2}^N$, either restricted to the case non-vanishing circulation
\begin{equation}
\sum_{i=1}^N\Gamma_i \neq 0\\
\end{equation}
or restricted to the part of phase space with non-vanishing total momentum
\begin{equation}
\sum_{i=1}^N \Gamma_i \mathbf r_i \neq 0.
\end{equation}
\begin{itemize}
\item If $N=4$, there exist configurations of point-vortices for which the motion is non-integrable.
\item If $N\geq 5$,  there exist configurations of point-vortices for which the Arnold diffusion occurs.
\end{itemize}
\end{thm}
Regarding the flat torus, \citet{KiAr2018} gave numerical evidence of non-integrability for three point-vortices on $\Tt^2$ with non-zero total circulation.
Our point-vortex gallery in \autoref{sec:gallery} also indicate chaotic behaviour in this case.
For hyperbolic space, we have also conducted numerical simulations in \autoref{sec:gallery} indicating the same behaviour as for the sphere: chaotic behaviour for four point-vortices with non-vanishing momentum.
To the best of our knowledge there are yet no rigorous non-integrability results available for the torus or the hyperbolic space.

\section{Outlook: long-time predictions for 2D Euler equations}\label{sec:outlook}

The prevailing theories for the long-time behaviour of the Euler equations~\eqref{eq:euler} on a 2-dimensional manifold are those given by \citet{Mi1990} and by \citet{RoSo1991}, here referred to as \emph{MRS}.
The MRS approach is a generalization of Onsager's~\cite{On1949} ideas, from discrete to continuous vorticity fields. 
These theories state that, in the long-time limit, the vorticity field evolves towards a state where the entropy of a course-grain probability distribution of macroscopic states is maximized under the constraint of conservation of energy and Casimirs.
Consequently, this leads to a course-grain steady vorticity state, characterized by a functional dependence between vorticity and stream function.
For a survey of MRS and the statistical approach to 2D turbulence, see the article by \citet{BoVe2012}.

However, in a numerical study for 2D Euler equations on the sphere, \citet{DrQiMa2015} gave results that contradict MRS theory, yielding, for randomly generated initial conditions, a seemingly non-steady vortex blob configuration.
The numerical method used in \cite{DrQiMa2015} did not conserve Casimir functions, which raised questions of the reliability since MRS theory is based on conservation of Casimirs.
As a remedy, we develop in~\cite{MoVi2020} a Casimir preserving numerical method for Euler equations on the sphere that captures all the features of 2D Euler equations: conservation of Casimirs, energy, and the Lie--Poisson structure. 
Using this method we obtain, again, strong evidence against the MRS predictions, but now with a more reliable, structure preserving method.
Furthermore, we found a new mechanism that connects the long-time behaviour with point-vortex integrability results.
Here is a brief outline of how it works:
\begin{enumerate}
	\item As is well-known in 2D turbulence, the \emph{inverse energy cascade}, discovered by \citet{Kr1969}, forces smaller vortex formations of the same sign to merge into larger ones by vortex stretching, forming positive and negative \emph{vortex blobs}.

	\item As long as the vortex blobs are not too close to each other, so they are not ripped apart and merged, their dynamics are well described by point-vortex dynamics. 
	Theoretical results in this direction are given, for example, in the monograph by \citet{MaPu2012}.

	\item The vortex merging continue, into fewer and larger vortex blobs, until the blob dynamics become integrable, with well separated vortex blob trajectories.
	Because of quasi-periodicity, the vortex blobs are then `stuck' in this part of phase space and no further merging occurs.

	\item A prediction for the final number of vortex blobs $N$ is thus given by integrability results for point-vortex dynamics: for the given fluid configuration (in terms of circulation, energy, momentum, etc.), find the largest $N$ such that the dynamics is integrable for $N$ vortices, but non-integrable for $N+1$ vortices.
\end{enumerate}
For the Euler equations on the sphere, Casimir preserving numerical simulations with randomly generated initial conditions, presented in \cite{MoVi2020}, perfectly align with this mechanism: if the total linear momentum is zero (or very small) we see the formation of 4 vortex blobs interacting in a quasi-periodic, non-steady fashion reflecting the results in Theorem~\ref{thm:PV_int_S2_zero_mom} above.
If the linear momentum is non-zero, we see the formation of 3 vortex blobs, reflecting Theorem~\ref{thm:PV_int_S2_non_zero_mom}.
We anticipate that the long-time behaviour on other domains also shall be reflected in the corresponding point-vortex integrability results, at least for the plane and the hyperbolic plane.
(As we can see in the proofs above, the flat torus is, from the symplectic reduction viewpoint, more complicated than the other cases; the connection to the corresponding Euler equations is not direct.)

But of course, numerical simulations alone, even if they preserve all the underlying structures, are not enough and must be accompanied with rigorous mathematical analysis, attempting to prove the connection between integrability and the long-time behaviour.
We consider the paper at hand the first step in this direction.

\medskip

\noindent\textbf{Acknowledgements.} 
The authors acknowledge support from the European Union Horizon 2020 research and innovation programmes under the Marie Skodowska-Curie grant agreement No.\ 691070, by the Swedish Foundation for International Cooperation in Research and Higher Eduction (STINT) grant No.\ PT2014-5823, by the Swedish Research Council (VR) grant No.\ 2017-05040,
and by the Knut and Alice Wallenberg Foundation (KAW) grant No.\ WAF2019.0201.

\appendix
\section{Gallery of point-vortex solutions}\label{sec:gallery}

\begin{figure}
\centering
\subfloat[3 point-vortices, vanishing momentum. Solutions are relative equilibria.]{
  \includegraphics[width=0.5\textwidth]{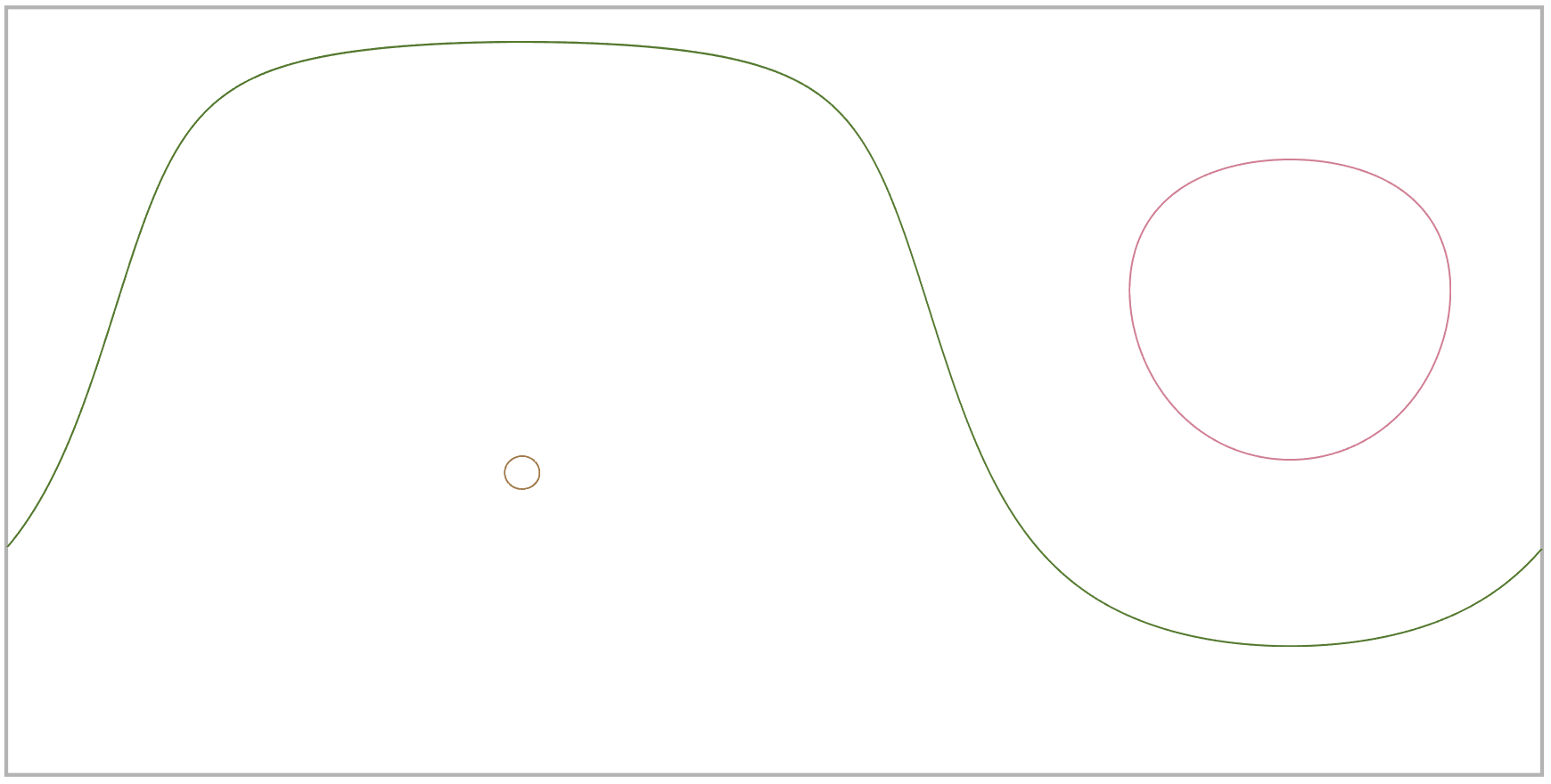}
}
\subfloat[3 point-vortices, non-vanishing momentum. Solutions are integrable.]{
  \includegraphics[width=0.5\textwidth]{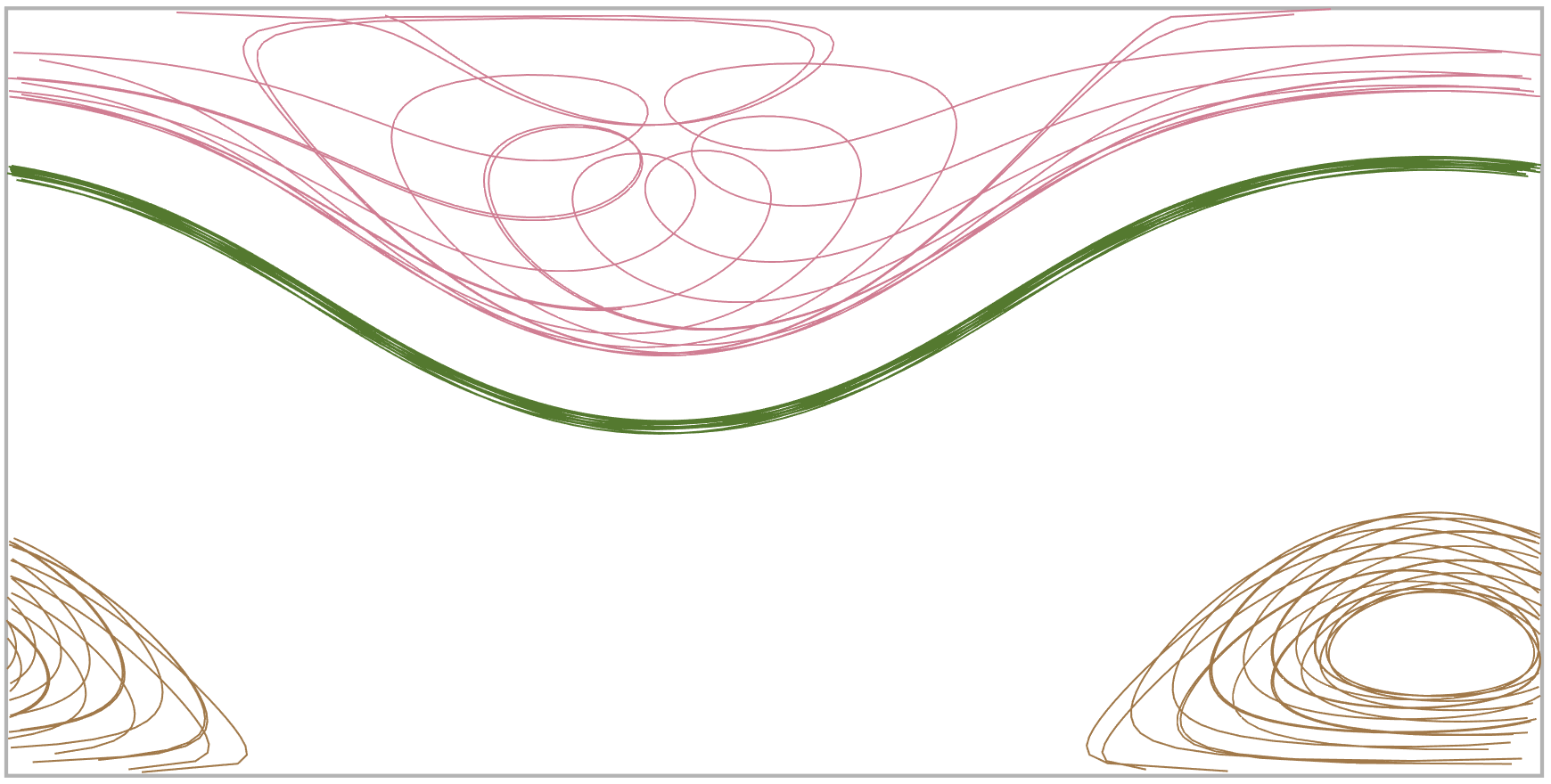}
}
\hspace{0mm}
\subfloat[4 point-vortices, vanishing momentum. Solutions are integrable.]{
  \includegraphics[width=0.5\textwidth]{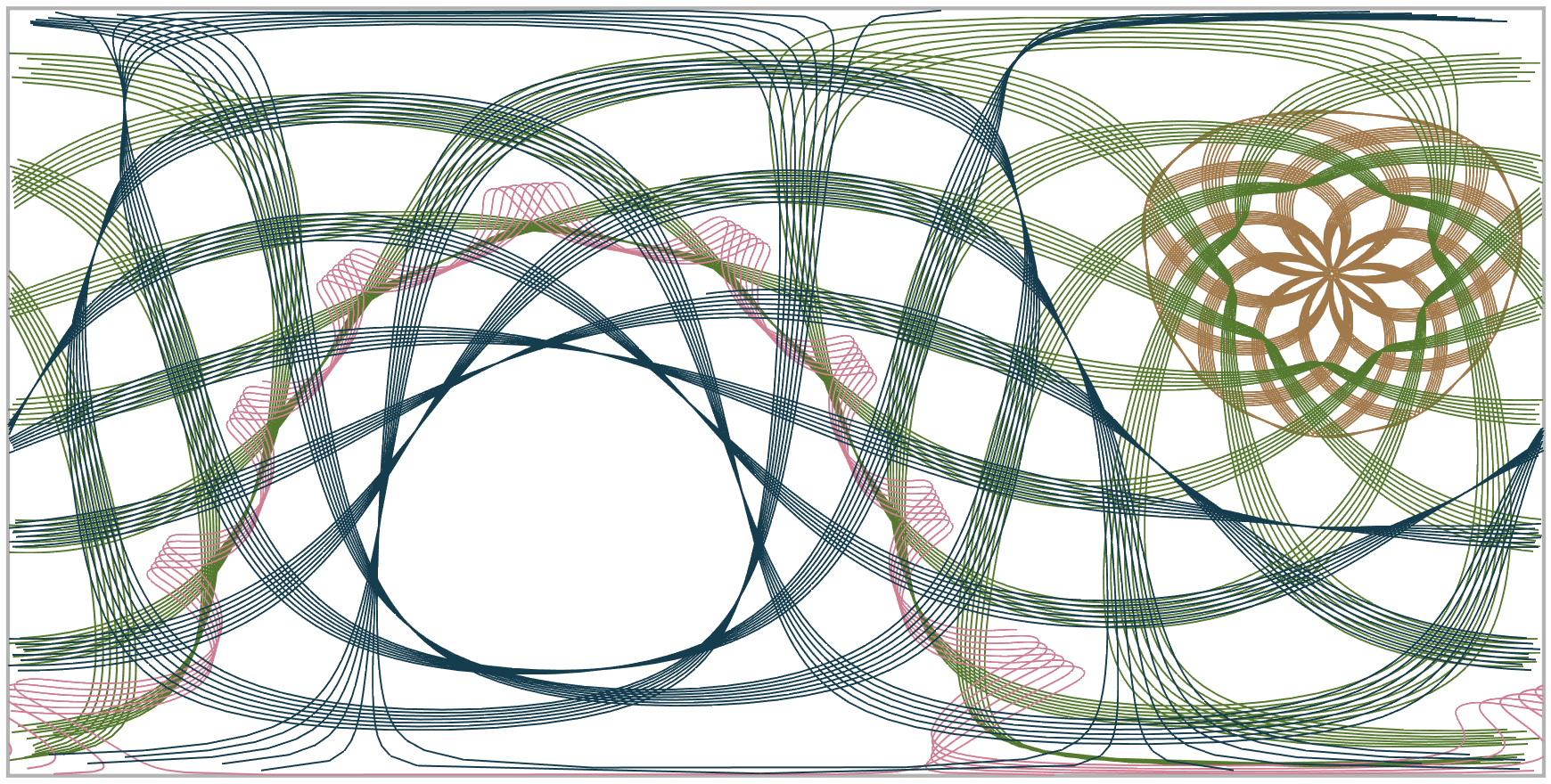}
}
\subfloat[4 point-vortices, non-vanishing momentum. Chaotic behaviour.]{
  \includegraphics[width=0.5\textwidth]{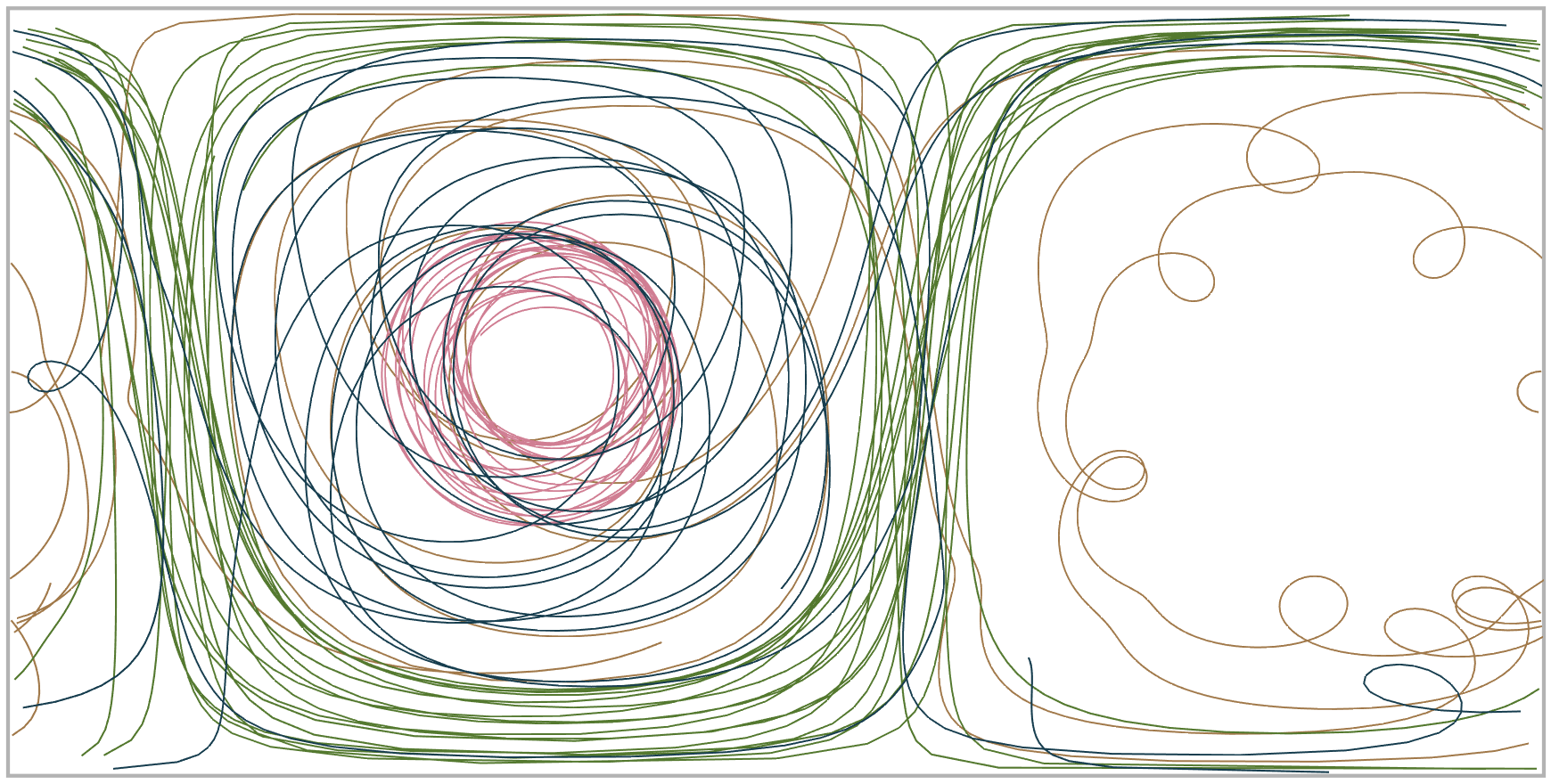}
}
\caption{\!\!\textsc{(sphere)} Trajectories of typical 3 or 4 point-vortex solutions on the sphere, for vanishing and non-vanishing momentum.
Each color represents one point-vortex trajectory. 
The trajectories are visualized using spherical coordinates.
The results align with the results in Theorem~\ref{thm:PV_int_S2_zero_mom} and Theorem~\ref{thm:PV_int_S2_non_zero_mom}; 
the trajectories are quasi-periodic in the situations where symplectic reduction can be used to prove integrability, and they are chaotic whenever symplectic reduction does not predict integrability. 
}\label{fig:sphere}
\end{figure}

In this appendix we showcase typical point-vortex phase space trajectories for all of the investigated geometries: the sphere, the plane, the hyperbolic plane, and the flat torus.
The results verify the integrability results in the theorems above: if conditions for integrability (or relative equilibria) are fulfilled, then beautiful, quasi-periodic patterns emerge.
Interestingly, our gallery of point-vortex trajectories also indicate that the symplectic reduction approach to integrability is sharp in the following sense: in all tested cases where symplectic reduction could not be used to prove integrability, the resulting trajectories appear to be chaotic.
This leads us to a conjecture about non-integrability.
Before that, we explain how to compute the trajectories and we give some specific comments on the results.

Initial conditions are generated randomly; for the compact domains they are drawn from the uniform distribution (with respect to the underlying Riemannian volume form).
For the plane, we use the multivariate Gaussian distribution with zero mean and unit variance.
The vortex strenghts are randomly generated from independent normal distributions.
For initial conditions with constraints, such as vanishing momentum or circulation, we use orthogonal projection.

To compute the trajectories we use numerical integration methods.
For the sphere we use the \emph{spherical midpoint method} \cite{McMoVe2014a,McMoVe2016b}.
For the flat geometries we use the \emph{implicit midpoint method} (cf.\ \cite{HaLuWa2006}).
For the hyperbolic plane we use the \emph{hyperbolic midpoint method} \cite{Vi2019} (see also \cite{ModViv2019a}). 
All these methods are second order approximations of the exact flow map.

Of course, the numerical discretizations introduce errors, so the computed trajectories are not true point-vortex trajectories.
One can argue that they therefore might not display the correct qualitative behaviour (such as quasi-periodic versus chaotic behaviour).
For general numerical methods this is a sound consideration.
Here, however, the numerical methods we use are specifically chosen to uphold the qualtative feature.
Indeed, all the methods we use are symplectic (with respect to the underlying symplectic structure) and furthermore equivariant (with respect to the underlying invariance group).
Since the methods are symplectic, one can utilize the framework of \emph{backward error analysis} (cf.~\cite{HaLuWa2006}) to prove that the discretized flow corresponds, for exponentially long time intervalls, to the exact flow of a slightly modified Hamiltonian system.
Furthermore, since the methods are equivariant, the modified Hamiltonian function is going to be invariant.
Thus, since the integrability Theorems~\ref{thm:PV_int_S2_non_zero_mom}-\ref{thm:PV_int_T2_zero_circ} above are valid not just for the Hamiltonian corresponding to point-vortex dynamics, but for any invariant Hamiltonian, we conclude that the theorems are valid also for the modified Hamiltonian systems yielding the computed trajectories, independent of the magnitude of the numerical errors.

\begin{figure}
\centering
\subfloat[3 point-vortices. Solutions are integrable.]{
  \includegraphics[width=0.5\textwidth]{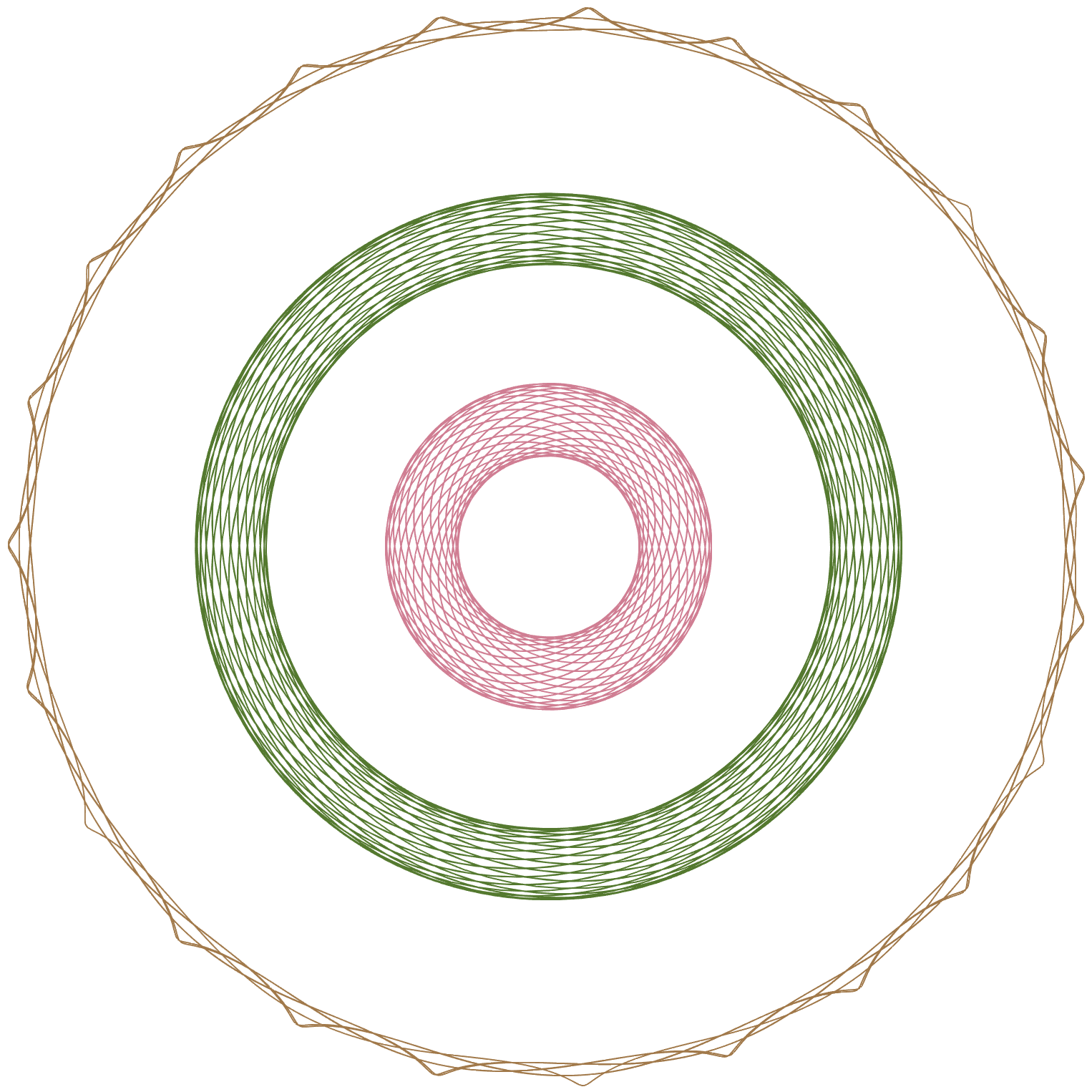}
}
\subfloat[4 point-vortices, vanishing momentum and circulation. Solutions are integrable.]{
  \includegraphics[width=0.5\textwidth]{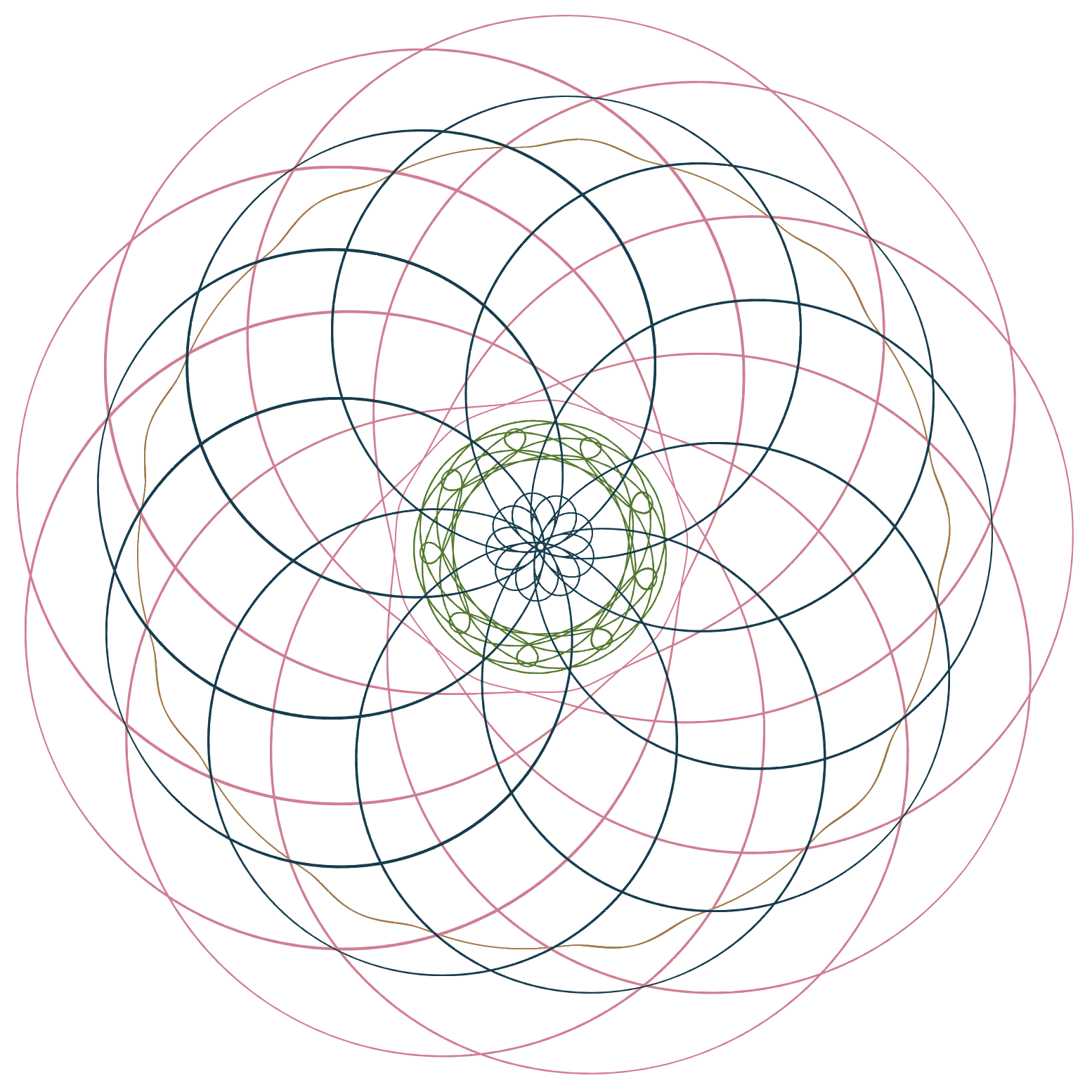}
}
\hspace{0mm}
\subfloat[4 point-vortices, non-vanishing momentum and vanishing circulation. Chaotic behaviour.]{
  \includegraphics[width=0.5\textwidth]{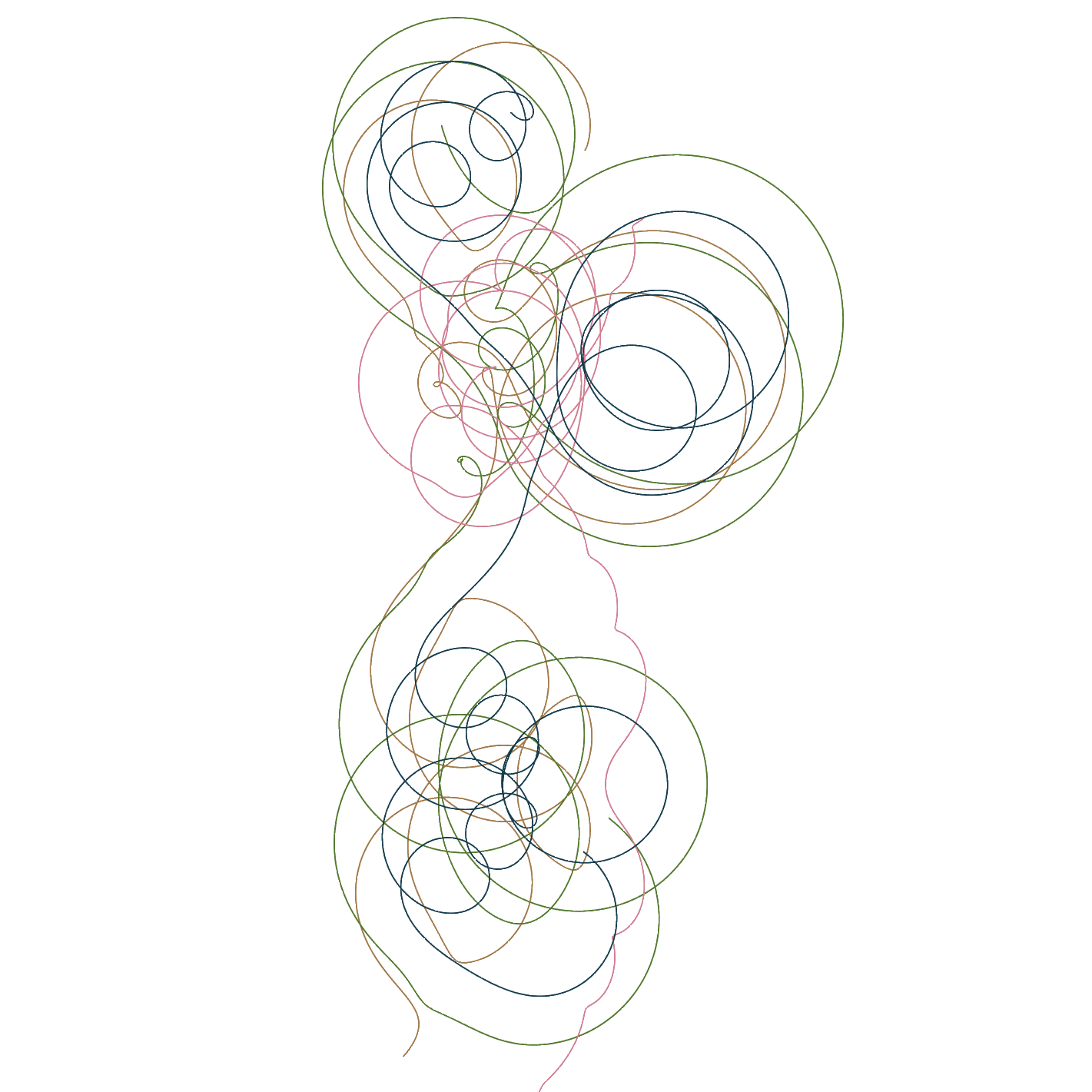}
}
\subfloat[4 point-vortices, vanishing momentum and non-vanishing circulation. Chaotic behaviour.]{
  \includegraphics[width=0.5\textwidth]{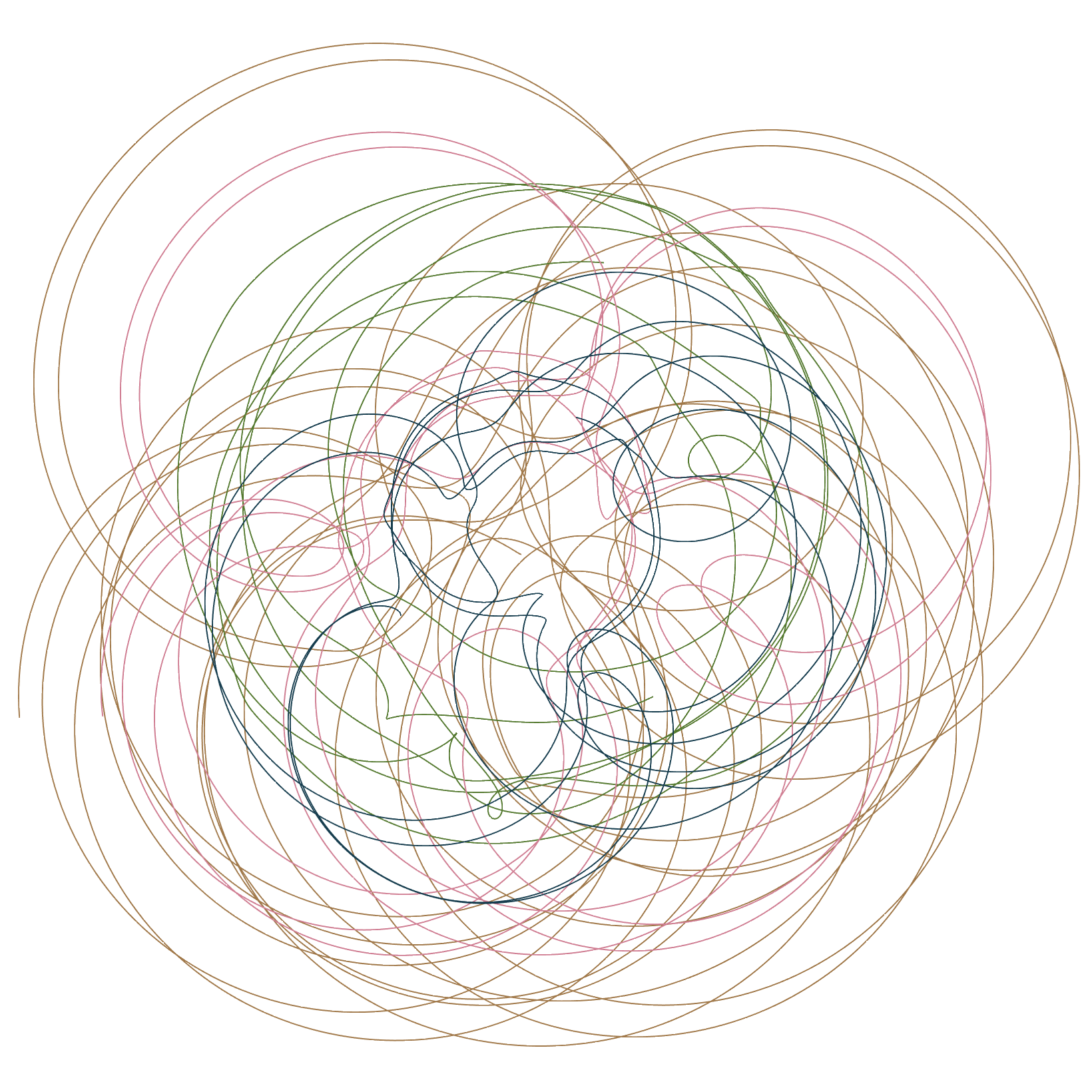}
}
\caption{\!\!\textsc{(plane)} Trajectories of typical 3 or 4 point-vortex solutions on the plane, for vanishing and non-vanishing momentum and circulation.
Each color represents one point-vortex trajectory. 
The results align with the results in Theorem~\ref{thm:PV_int_plane_zero_circ_zero_mom} and Theorem~\ref{thm:PV_int_plane_non_zero_circ_zero_mom}; 
the trajectories are quasi-periodic in the situations where symplectic reduction can be used to prove integrability, and they are chaotic whenever symplectic reduction does not predict integrability. 
}\label{fig:plane}
\end{figure}

\begin{figure}
\centering
\subfloat[3 point-vortices, vanishing momentum. Solutions are relative equilibria.]{
  \includegraphics[width=0.5\textwidth]{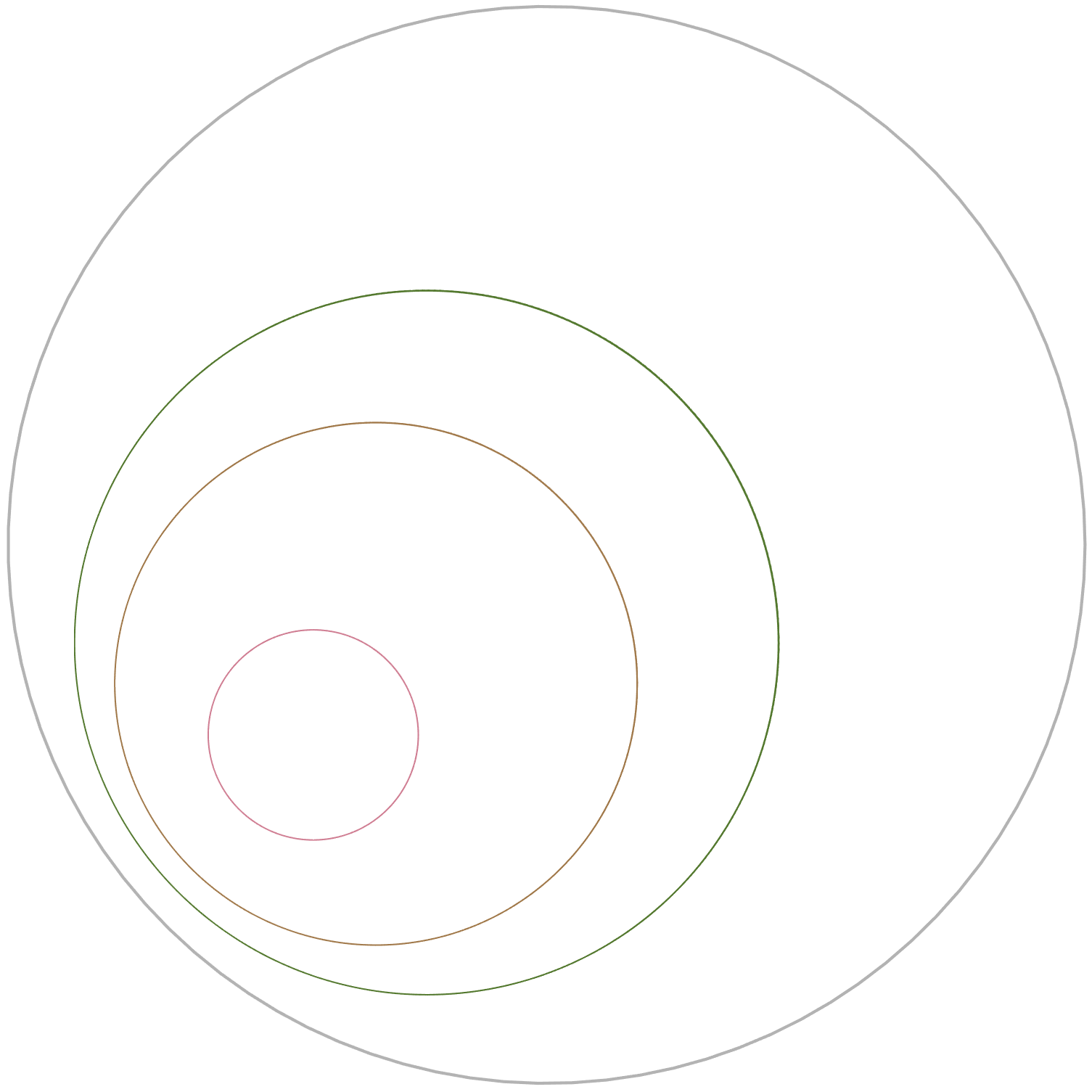}
}
\subfloat[3 point-vortices, non-vanishing momentum. Solutions are integrable.]{
  \includegraphics[width=0.5\textwidth]{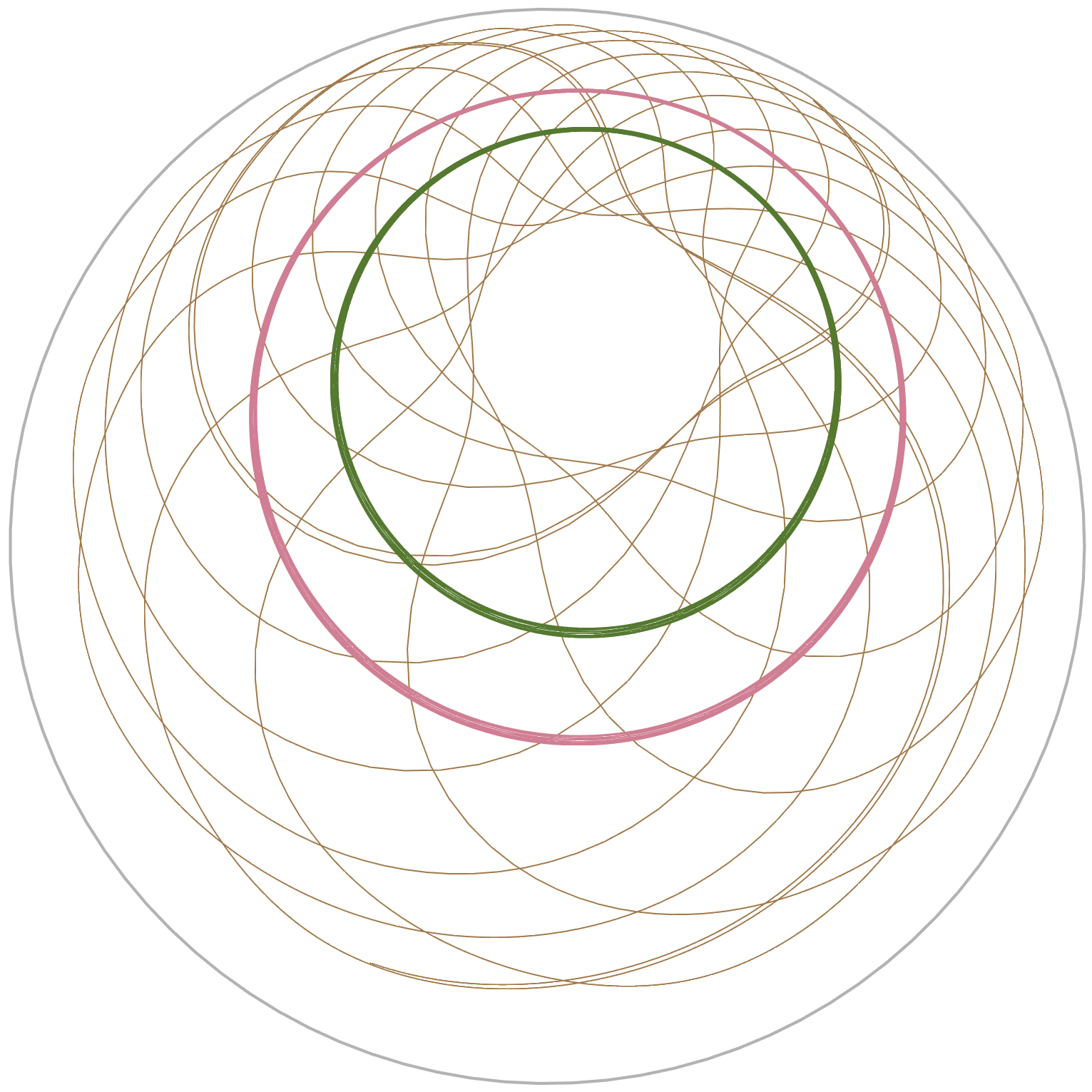}
}
\hspace{0mm}
\subfloat[4 point-vortices, vanishing momentum. Solutions are integrable.]{
  \includegraphics[width=0.5\textwidth]{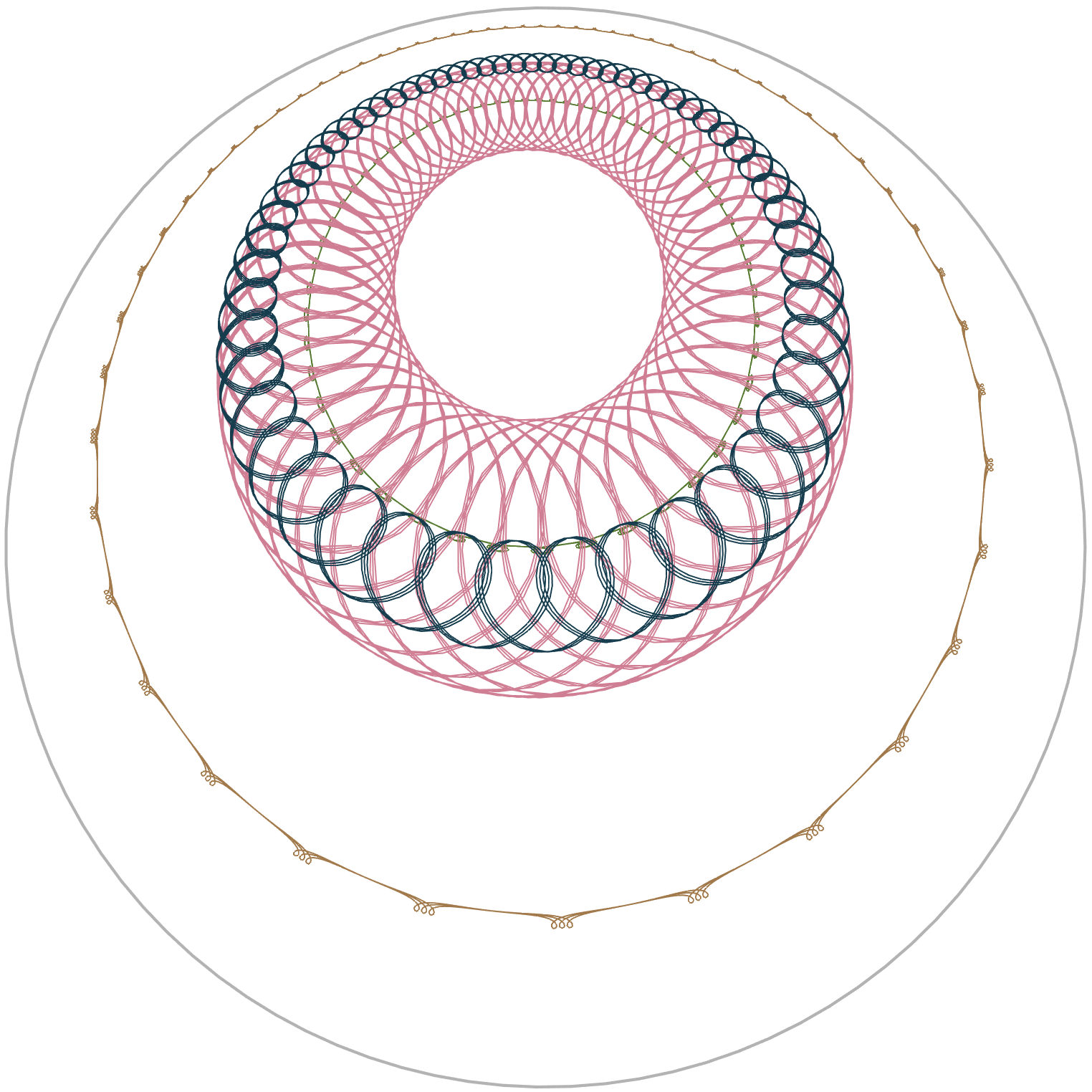}
}
\subfloat[4 point-vortices, non-vanishing momentum. Chaotic behaviour.]{
  \includegraphics[width=0.5\textwidth]{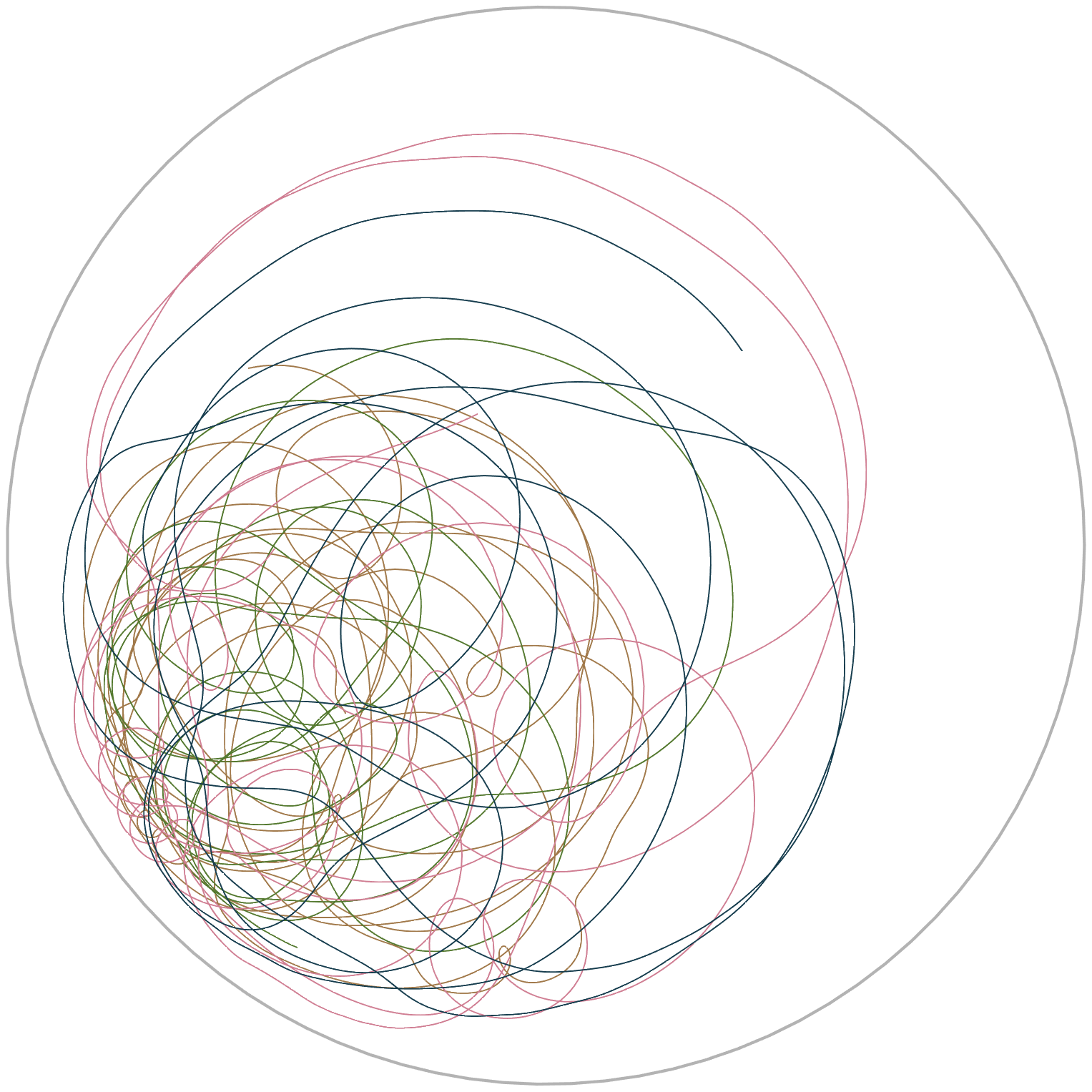} 
}
\caption{\!\!\textsc{(hyperbolic plane)} Trajectories of typical 3 or 4 point-vortex solutions on the hyperbolic plane, for vanishing and non-vanishing momentum and circulation.
The trajectories are visualized using the Poincaré disk model.
Each color represents one point-vortex trajectory. 
The results align with the results in Theorem~\ref{thm:PV_int_hyp_zero_mom} and Theorem~\ref{thm:PV_int_hyp_non_zero_mom}; 
the trajectories are quasi-periodic in the situations where symplectic reduction can be used to prove integrability, and they are chaotic whenever symplectic reduction does not predict integrability.
}\label{fig:hyplane}
\end{figure}

\begin{figure}
\centering
\subfloat[2 point-vortices, vanishing circulation. Solutions are relative equilibria.]{
  \includegraphics[width=0.5\textwidth]{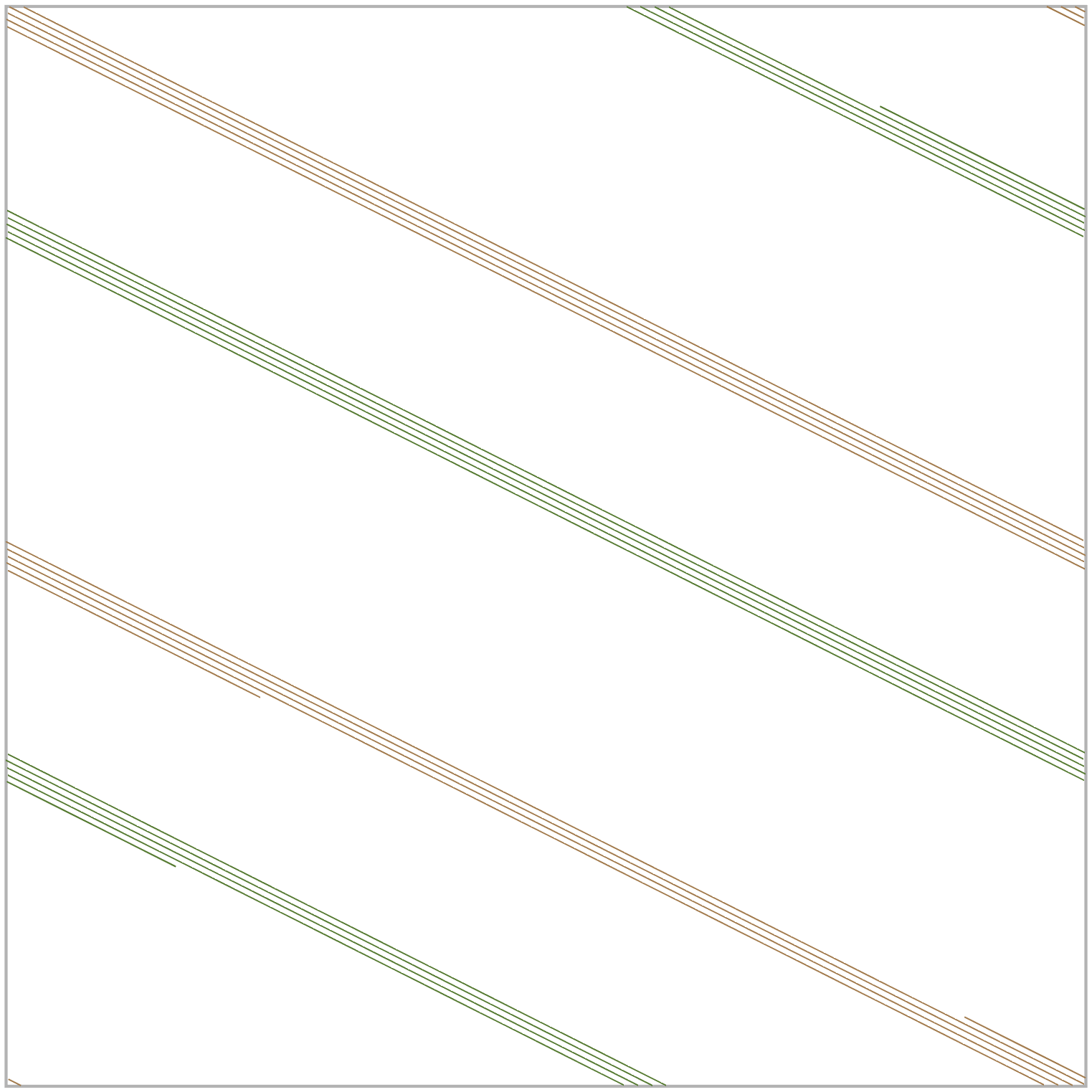}
}
\subfloat[2 point-vortices, non-vanishing circulation. Solutions are integrable.]{
  \includegraphics[width=0.5\textwidth]{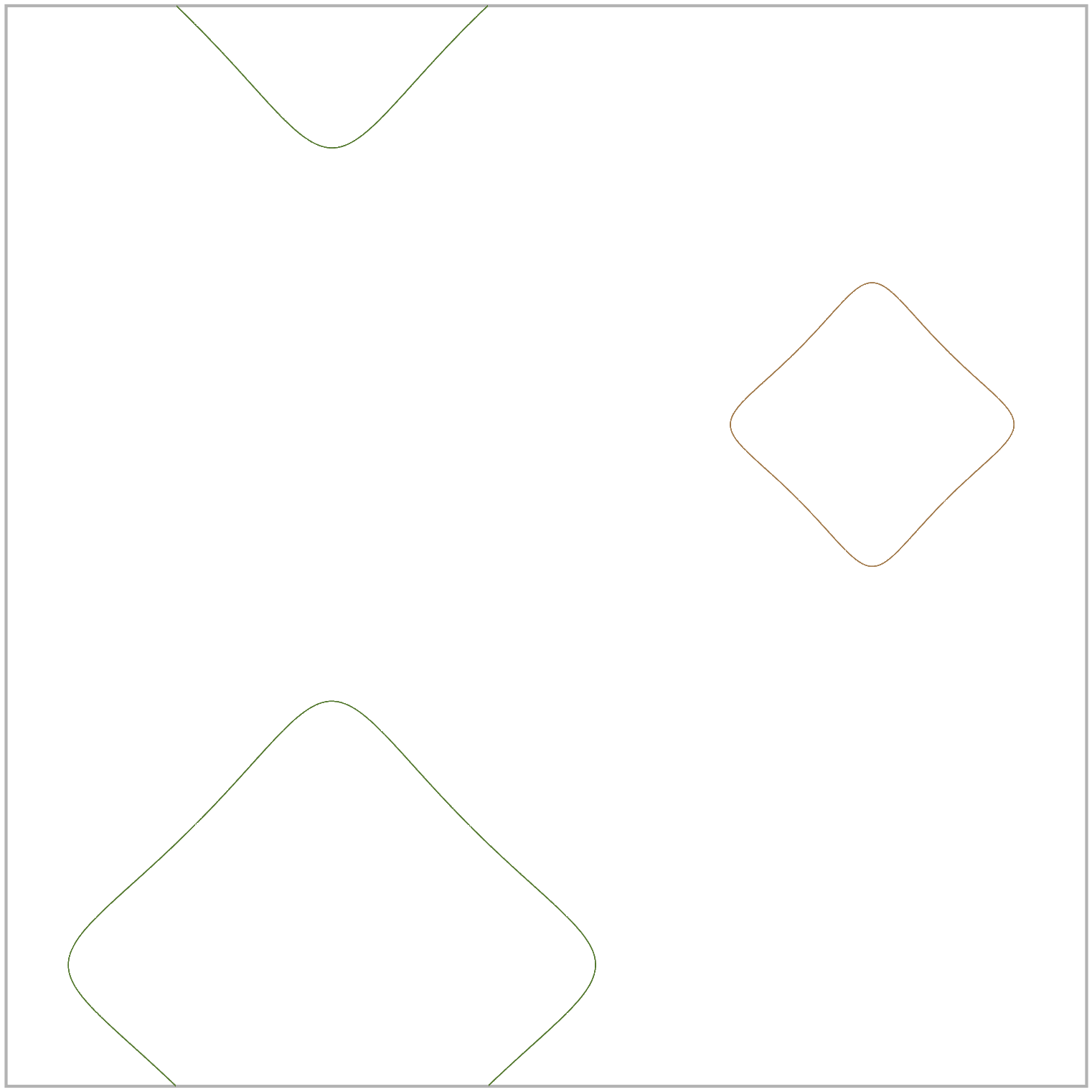}
}
\hspace{0mm}
\subfloat[3 point-vortices, vanishing circulation. Solutions are integrable.]{
  \includegraphics[width=0.5\textwidth]{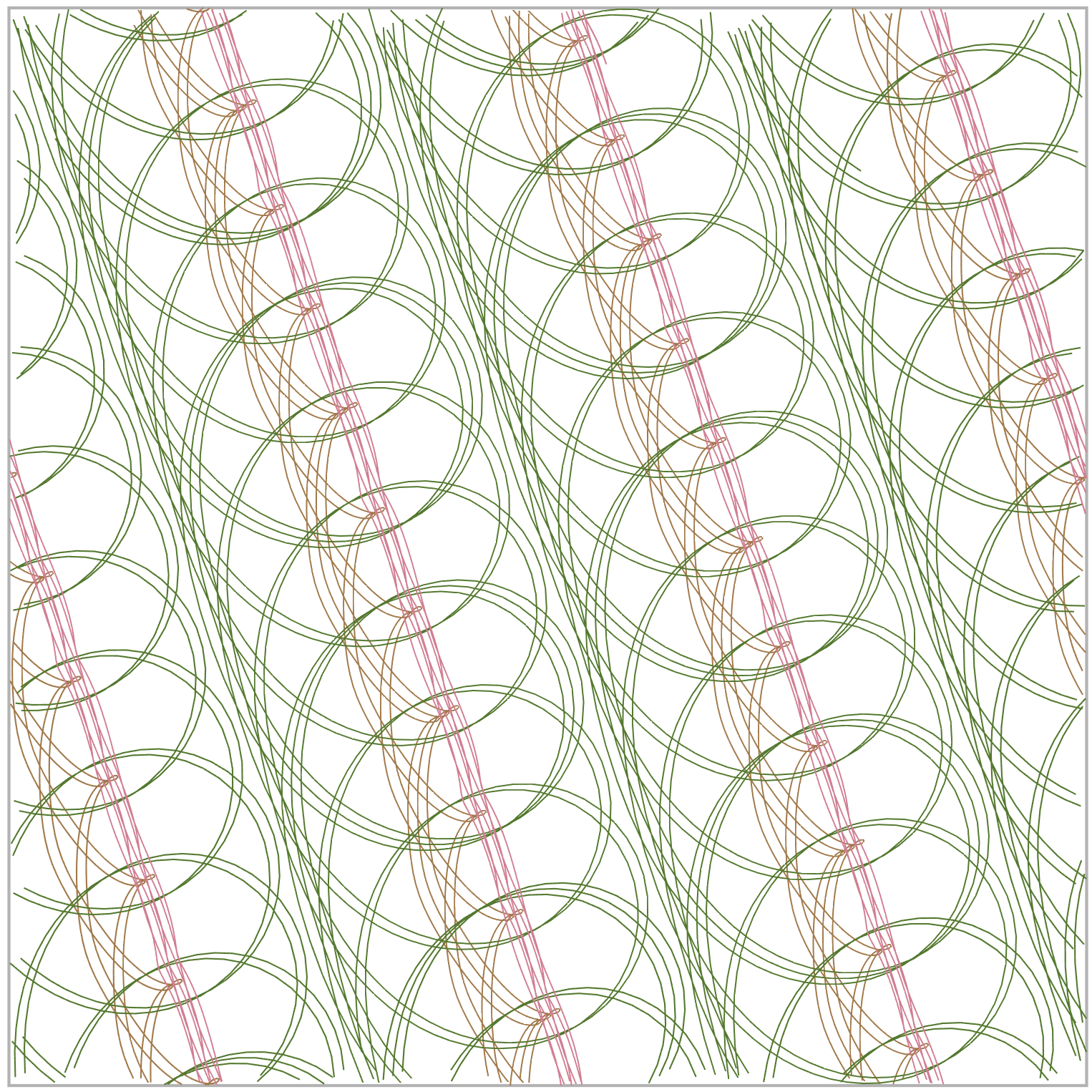}
}
\subfloat[3 point-vortices, non-vanishing circulation. Chaotic behaviour.]{
  \includegraphics[width=0.5\textwidth]{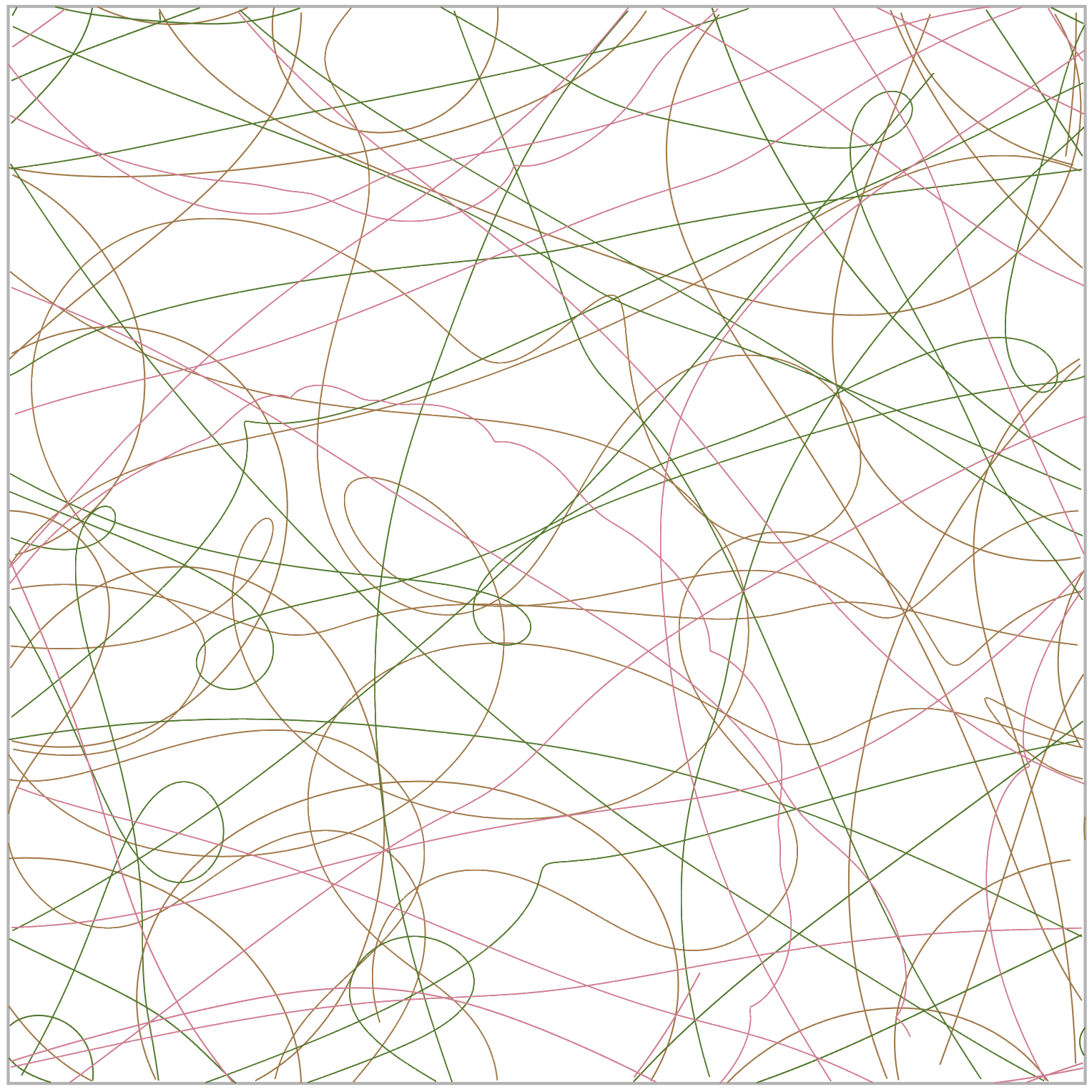} 
}
\caption{\!\!\textsc{(flat torus)} Trajectories of typical 2 or 3 point-vortex solutions on the flat torus, for vanishing and non-vanishing circulation.
Each color represents one point-vortex trajectory. 
Notice that solution trajectories are doubly periodic over the boundaries.
The results align with the results in Theorem~\ref{thm:PV_int_T2_zero_circ} and Theorem~\ref{thm:PV_int_T2_non_zero_circ}; 
the trajectories are quasi-periodic in the situations where symplectic reduction can be used to prove integrability, and they are chaotic whenever symplectic reduction does not predict integrability.
}\label{fig:torus}
\end{figure}

For each of the settings in the theorems (vanishing momentum, non-vanishing momentum, etc.), we generated several trajectories for each geometry.
To check the sharpness of our integrability results, we also computed solutions for the simplest possible setting for which the appropriate theorem do not apply.
The typical behaviour is illustrated in Figures~\ref{fig:sphere}-\ref{fig:torus}; this is our gallery of point vortex solutions.

For the sphere, in \autoref{fig:sphere}, we obtain quasi-periodic or relative equilibria solution in accordance with Theorems~\ref{thm:PV_int_S2_non_zero_mom}-\ref{thm:PV_int_S2_zero_mom}.
For 4 point-vortices with non-vanishing momentum the behaviour is chaotic; a typical example is shown in \autoref{fig:sphere}(D).
Also for the plane, in \autoref{fig:plane}, we see quasi-periodic solutions whenever Theorems~\ref{thm:PV_int_plane_zero_circ_zero_mom}-\ref{thm:PV_int_plane_non_zero_circ_zero_mom} apply.
Recall that integrability for the planar 4 point-vortex solutions depend on two conditions: both momentum and circulation have to vanish.
If either of these conditions are not fulfilled, the trajectories become chaotic; typical examples are shown in \autoref{fig:plane}(C)-(D).
Interestingly, from the point-of-view of symplectic reduction, the non-vanishing of momentum and circulation corresponds to two different violations of the conditions needed: non-vanishing momentum implies that the symmetry group is too small to obtain integrability, whereas non-vanishing circulation implies that the momentum map is not equivariant (which is a condition in Theorem~\ref{thm:Symplectic Reduction theorem}).
Similarly, in \autoref{fig:hyplane} for the hyperbolic plane, and in \autoref{fig:torus} for the flat torus, we observe the behaviour predicted by the corresponding integrability theorem, and whenever the conditions are not fulfilled we observe chaotic trajectories.
This leads us to formulate the following:

\begin{conj}
	Let $M$ be a 2-dimensional Riemannian manifold with symmetry group $G$.
	Then a generic $G$-invariant Hamiltonian system on $P^N_M$ is integrable only when the geometric conditions (free and proper action, and equivariant momentum map) are such that Theorem~\ref{thm:Symplectic Reduction theorem} directly implies integrability.
\end{conj}

\bibliographystyle{amsplainnat}
\bibliography{biblio}
 \end{document}